\documentclass[
  journal=pasa,
  manuscript=research-paper, %% or "review"
  year=2022,
  volume=??,
]{cup-journal}

\DeclareUnicodeCharacter{2212}{-}
\newcommand{\frb}{FRB\,20171020A} % from TNS

\newcommand{\eso}{ESO~601−G036}

\newcommand{\ppcc}{pc cm$^{-3}$}
\newcommand{\wise}{WISE $\times$ SCOSPZ}
\newcommand{\hipass}{H\,{\sc i} Parkes All Sky Survey}
\newcommand{\HI}{H\,{\sc i}}
\def\lesssim{\mathrel{\hbox{\rlap{\hbox{%
 \lower4pt\hbox{$\sim$}}}\hbox{$<$}}}}
\def\gtrsim{\mathrel{\hbox{\rlap{\hbox{%
 \lower4pt\hbox{$\sim$}}}\hbox{$>$}}}}

\def\arcmin{\hbox{$^\prime$}}
\def\arcsec{\hbox{$^{\prime\prime}$}}

\def\farcs{\hbox{$.\!\!^{\prime\prime}$}}

\usepackage{microtype,siunitx,booktabs,lipsum,amsmath,url,xcolor}

\title{The Host Galaxy of \frb\ Revisited}

\author{Karen Lee-Waddell}
\affiliation{International Centre for Radio Astronomy Research, The University of Western Australia, 35 Stirling Highway, Crawley WA 6009, Australia}
\alsoaffiliation{CSIRO Space and Astronomy, P.O. Box 1130, Bentley WA 6102, Australia}
\alsoaffiliation{International Centre for Radio Astronomy Research, Curtin University, Bentley, WA 6102, Australia}
\author{Clancy W.\ James}
\affiliation{International Centre for Radio Astronomy Research, Curtin University, Bentley, WA 6102, Australia}
\author{Stuart D. Ryder}
\affiliation{School of Mathematical and Physical Sciences, Macquarie University, NSW 2109, Australia}
\alsoaffiliation{Astronomy, Astrophysics and Astrophotonics Research Centre, Macquarie University, Sydney, NSW 2109, Australia}
\author{Elizabeth K. Mahony}
\affiliation{CSIRO Space and Astronomy, Australia Telescope National Facility, P.O. Box 76, Epping, NSW 1710, Australia}

\author{Arash Bahramian}
\affiliation{International Centre for Radio Astronomy Research, Curtin University, Bentley, WA 6102, Australia}

\author{B\"arbel S. Koribalski}
\affiliation{CSIRO Space and Astronomy, Australia Telescope National Facility, P.O. Box 76, Epping, NSW 1710, Australia}
\alsoaffiliation{School of Science, Western Sydney University, Locked Bag 1797, Penrith, NSW 2751, Australia}
\author{Pravir Kumar}
\affiliation{Centre for Astrophysics and Supercomputing, Swinburne University of Technology, P.O. Box 218, Hawthorn, VIC 3122, Australia}

\author{Lachlan Marnoch}
\affiliation{School of Mathematical and Physical Sciences, Macquarie University, NSW 2109, Australia}
\alsoaffiliation{Astronomy, Astrophysics and Astrophotonics Research Centre, Macquarie University, Sydney, NSW 2109, Australia}
\alsoaffiliation{CSIRO Space and Astronomy, Australia Telescope National Facility, P.O. Box 76, Epping, NSW 1710, Australia}
\alsoaffiliation{ARC Centre of Excellence for All Sky Astrophysics in 3 Dimensions (ASTRO 3D)}

\author{Freya O. North-Hickey}
\affiliation{International Centre for Radio Astronomy Research, Curtin University, Bentley, WA 6102, Australia}

\author{Elaine M. Sadler}
\affiliation{Sydney Institute for Astronomy, School of Physics A28, University of Sydney, NSW 2006, Australia}
\alsoaffiliation{CSIRO Space and Astronomy, Australia Telescope National Facility, P.O. Box 76, Epping, NSW 1710, Australia}
\alsoaffiliation{ARC Centre of Excellence for All Sky Astrophysics in 3 Dimensions (ASTRO 3D)}

\author{Ryan Shannon}
\affiliation{Centre for Astrophysics and Supercomputing, Swinburne University of Technology, P.O. Box 218, Hawthorn, VIC 3122, Australia}
\alsoaffiliation{ARC Centre of Excellence for Gravitational Wave Discovery (OzGrav)}

\author{Nicolas Tejos}
\affiliation{Instituto de F\'isica, Pontificia Universidad Cat\'olica de Valpara\'iso, Casilla 4059, Valpara\'iso, Chile}

\author{Jessica E. Thorne}
\affiliation{International Centre for Radio Astronomy Research, The University of Western Australia, 35 Stirling Highway, Crawley WA 6009, Australia}

\author{Jing Wang}
\affiliation{Kavli Institute for Astronomy and Astrophysics, Peking University, Beijing 100871, China}

\author{Randall Wayth}
\affiliation{International Centre for Radio Astronomy Research, Curtin University, Bentley, WA 6102, Australia}

\received {20 Feb 2023}
\revised  {10 May 2023}
\accepted {25 May 2023}
\published{dd Mmm YYYY}


\begin{document}

\begin{abstract}
The putative host galaxy of \frb\ was first identified as \eso\ in 2018, but as no repeat bursts have been detected, direct confirmation of the host remains elusive. In light of recent developments in the field, we re-examine this host and determine a new association confidence level of 98\%. At 37\,Mpc, this makes \eso\ the third closest FRB host galaxy to be identified to date and the closest to host an apparently non-repeating FRB (with an estimated repetition rate limit of $<0.011$\,bursts per day above $10^{39}$\,erg).
Due to its close distance, we are able to perform detailed multi-wavelength analysis on the \eso\ system. Follow-up observations confirm \eso\ to be a typical star-forming galaxy with \HI\ and stellar masses of $\log_{10}(M_{\mbox{\HI}} / M_\odot) \sim 9.2$ and $\log_{10}(M_\star / M_\odot) = 8.64^{+0.03}_{-0.15}$, and a star formation rate of $\text{SFR} = 0.09 \pm 0.01\,M_\odot\,\text{yr}^{-1}$. 
We detect, for the first time, a diffuse gaseous tail ($\log_{10}(M_{\mbox{\HI}} / M_\odot) \sim 8.3$) extending to the south-west that suggests recent interactions, likely with the confirmed nearby companion ESO~601-G037. ESO~601-G037 is a stellar shred located to the south of \eso\ that has an arc-like morphology, is about an order of magnitude less massive, and has a lower gas metallicity that is indicative of a younger stellar population. The properties of the \eso\ system indicate an ongoing minor merger event, which is affecting the overall gaseous component of the system and the stars within ESO~601-G037. Such activity is consistent with current FRB progenitor models involving magnetars and the signs of recent interactions in other nearby FRB host galaxies. 

\end{abstract}

\section{Introduction} \label{sec:intro}

Fast radio bursts (FRBs) are currently one of the most tantalising mysteries in astrophysics as they are particularly peculiar events arising from unknown origins. FRBs manifest themselves as extremely luminous (brightness temperature $T_b \sim 10^{35}$ K) millisecond-duration radio pulses characterised by integrated electron column density contributions (as traced by the dispersion measure; DM) well in excess of the Galactic contribution along the line-of-sight \citep{thornton_population_2013}. Since their discovery by the Parkes 64m {\em Murriyang} radio-telescope by \citet{lorimer_bright_2007}, over 600 FRBs have been detected with numerous instruments \citep{Petroff2022}. With wide field-of-view and sensitive searches for FRBs ongoing, e.g.\ with the Australian Square Kilometer Array Pathfinder \citep[ASKAP; ][]{shannon_dispersionbrightness_2018}, the Canadian Hydrogen Intensity Mapping Experiment  \citep[CHIME; ][]{CHIME2021}, MeerKAT \citep{2022MNRAS.514.1961R}, and the Deep Synoptic Array \citep[DSA; ][]{2023arXiv230101000R}, there will be a significant increase in the number of detected FRB events in future years.

The $\sim 0.1-10$ ms duration of FRBs places upper limits on their emission regions to $<$ 30--3000~km in size, allowing for smearing due to scattering and detector time resolution. A multitude of progenitor models have been proposed to explain the mechanism(s) to produce FRBs \citep{Platts2019}. Leading models invoke highly magnetised neutron stars (i.e.\ magnetars) that result from the core-collapse of massive stars, which is strongly supported by the association of the Galactic FRB-like source FRB 20200428 with the magnetar SGR 1935+2154 \citep{Bochanek2020, CHIME2020}. The association of particularly active repeating FRBs with star-forming regions or persistent radio sources \citep{chatterjee_direct_2017,marcote_repeating_2020,Niu2022_190520B} is further evidence. The recent localisation of the repeating FRB~20200120E to a globular cluster in the M81 galactic system \citep{Kirsten2022} challenges this scenario; nevertheless, it has been hypothesised that magnetars can also form from the accretion-induced collapse of a white dwarf or merger of compact stars in a binary system \citep{duncan1992,Rosswog2003,Tauris2013}.

Localisation of FRBs to host galaxies is also crucial for investigating other potential progenitor avenues through exploration of global properties and the general host galaxy population. Studies thus far have identified a wide diversity in current FRB host galaxies suggesting that the broader population of FRBs can arise from both young and (moderately) old progenitors \citep{bhandari_host_2020, Heintz2020, Manning2021, Bhandari2022}. Detailed studies of these hosts, especially using multi-wavelength observations to fully characterise properties of the galaxy and its environment, would help in understanding the origin of FRBs. However, most FRB host galaxies are too distant to enable such analysis. Some nearby detections have provided useful upper limits on FRB counterparts at X-ray \citep{Mereghetti2021} or optical wavelengths \citep{Andreoni_2020}.

A small number of very nearby FRB host galaxies have been the subject of detailed radio wavelength investigations. Neutral hydrogen (\HI) analysis by \citet{Michalowski2021} of three galaxies --- NGC~3252, M81, and the Milky Way --- connects FRBs to recent enhancement of star formation due to galactic interactions. \citet{Kaur2022} come to a similar conclusion about the \HI-rich host galaxy of FRB~20180916B, which appears to be involved in a recent minor merger (when a low-mass satellite galaxy merges with its host galaxy). Looking at the molecular gas of the host of FRB~20180924B, \citet{Hsu2023} report a disturbed kinetic gas structure in the galaxy. In these aforementioned systems, all five FRB host galaxies have highly asymmetric emission line spectra, linking FRB progenitors to recent galaxy interaction events. However, recent work by \cite{Glowacki2023}, finds the \HI\ in the host galaxy of FRB~20211127I to be relatively undisturbed. With only a handful of current examples, detailed analysis of the gas in and around FRB host galaxies is still an emerging topic. As such, additions to the sample size are required to draw better conclusions about FRB progenitors.

\cite{shannon_dispersionbrightness_2018} presented the discovery of 20 FRBs in a fly's eye survey conducted at a Galactic latitude of $|b| = 50^{\circ} \pm 5^{\circ}$ carried out as part of the Commensal Real-time ASKAP Fast Transients \citep[CRAFT:][]{Macquart2010} survey on ASKAP. Here we focus on one event from that survey, \frb. 
Given the low DM of 114~\ppcc, \citet[][hereafter M18]{Mahony2018} found \eso\ --- an Sc galaxy located at RA 22:15:24.75, Dec -19:35:07.0 (J2000) --- to be the most probable host galaxy candidate given its low redshift, $z=0.00867$, corresponding to a central velocity of 2584 $\mathrm{km~s^{-1}}$ measured from \HI\ emission detected in the \hipass\ \citep[HIPASS;][]{Meyer2004} and its position close to the center of the error ellipse, according to ASKAP's beam. \cite{daCosta1998} found a comparable optical radial velocity for \eso\ of 2539 $\mathrm{km~s^{-1}}$. Using the \cite{Mould2000} model, the distance to this galaxy is 37~Mpc. \eso\ contains no compact, persistent radio continuum source while ultraviolet (UV) imaging from GALEX \citep{Martin2005} and spectroscopic observations reveal a low metallicity with a star formation rate (SFR) of $\mathrm{\sim 0.1~M_\odot~yr^{−1}}$ (M18). A nearby stellar shred, ESO 601-G037, can be detected in the GALEX images and was treated by M18 as part of the \eso\ system.

At the time of the M18 analysis, the only FRB localised to its host galaxy was the first known repeater, FRB~20121102A in a dwarf galaxy \citep{Tendulkar2017}. Since then, more than two dozen FRBs have been localised to arcsecond-level precision, the majority of these by CRAFT (Shannon et al. 2023, in prep.) and showing no repeat bursts. 

In this paper, we re-examine the case for \eso\ being the host galaxy of \frb\, in light of improvements in the localisation algorithm, and present detailed multi-wavelength follow-up observations of \eso\ that help to constrain the nature of FRB host galaxies in general. This analysis is especially pertinent since, despite the plethora of new discoveries, \eso\ is potentially the third closest FRB host galaxy to be identified, after M81 \citep{Bhardwaj2021_M81} and NGC~3252 \citep{Bhardwaj2021NGC} and the closest system to host a non-repeating FRB.
In Section \ref{sec:host_constraints}, we demonstrate our improved probabilistic association of this FRB to its host galaxy and further support this association by calculating the most probable redshift for this event from the FRB's properties. We present our multi-wavelength follow-up of the host galaxy in Section \ref{sec:follow-up}. In Section \ref{sec:discussion}, we discuss the implications of the host galaxy properties on FRB progenitor models. 

\section{New constraints on the host galaxy of \frb}
\label{sec:host_constraints}
\subsection{Improved localisation of \frb}\label{sec:loc}
The localisation region of \frb\ presented in M18 was based on signal-to-noise ratio (SNR) measurements of the detection in adjacent ASKAP beams. A Bayesian algorithm, with flat priors on Right Ascension and Declination, and a prior on fluence $P(F) \propto F^{-1}$ (i.e.\ flat in log-space), was used to find posterior probabilities for the FRB arrival direction as per the method in \citet{shannon_dispersionbrightness_2018}. The results were then fitted with an elliptical function to obtain characteristic localisation bounds.

\begin{figure}
    \centering
    \includegraphics[width=0.95\textwidth]{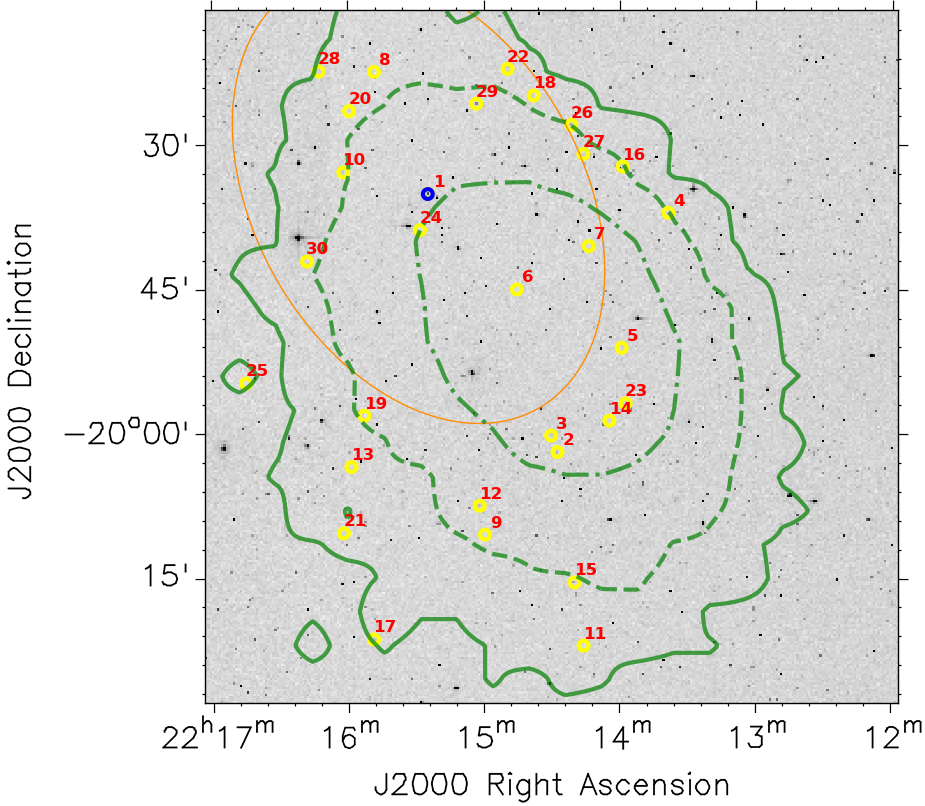}
    \caption{Full Bayesian posterior map with 1,2,3-sigma contours overlaid on a DSS2 Red image. The candidate galaxies (out to $z$ < 0.08) are circled in yellow and blue (where blue indicates \eso) and numbered indicating the most-likely host, in descending order, for \frb\ (refer to Table~\ref{tab:cands} for additional details). The $2 \sigma$ error ellipse from M18 is shown in orange for comparison.}
    \label{fig:lmap}
\end{figure}

Here we present two updates to this method. Firstly, the FRB fluence distribution is now known to closely follow a Euclidean distribution, $P(F) \propto F^{-5/2}$ \citep{james_slope_2019}, arguing for a steeper prior on $F$. However, such a prior is agnostic of DM, which is a proxy for distance, and the result is obtained by only averaging over distance. For FRBs constrained to a certain distance range (e.g.\ through DM cuts), the distribution of $F$ will approach that of the intrinsic FRB luminosity function, which is flatter --- at least until any cutoff due to a maximum FRB energy. Such a result has been found recently by the \citet{CHIME2021}. We therefore use a prior on $F$ of $P(F) \propto F^{-2.1}$, where $-2.1$ is the differential power-law slope of the FRB luminosity function from \citet{James2022Letter}. Secondly, we present results without the elliptical fit, i.e.\ using the full Bayesian posterior map, as shown in Figure~\ref{fig:lmap}. From this map, we define a localisation probability $p(s) = 1-{\rm C.L.}$, where C.L.\ is the confidence limit at which the FRB is found (thus an FRB on the $1 \sigma$ contour has $p(s) \approx 0.32$). 

The effect of our updated prior on $p(F)$ is that intrinsically dimmer FRBs are preferred, making it less likely that \frb\ was observed far from the beam centre. This shifts the localisation region some 20\arcmin\ to the south-west and emphasises the importance of the choice of prior in FRB localisations. The resulting 68\% confidence region is well-fit by an ellipse, centred at RA = $22^{\rm h} 14^{\rm m} 32^{\rm s}$, Dec = $-19^{\circ} 48^{\prime} 54^{\prime\prime}$ (J2000),  with semi-major axis $0.28^{\circ}$ oriented $35^{\circ}$ east from north, and semi-minor axis $0.2^{\circ}$.

We therefore update the list of candidate host galaxies from M18 to include all those within the $3 \sigma$ localisation and $z<0.08$ (the maximum redshift considered in M18). Using the NASA/IPAC Extragalactic Database (NED), we find five galaxies matching these criteria. Searching \wise\ \citep{WiseSuperCosmos}, which is deeper than NED and limited primarily by SuperCOSMOS completeness to optical magnitudes $R<19.5$ and $B<21$, there are 32 such galaxies, including five co-located with those listed in NED. Two of these, however, are located at small ``islands'' of probability almost one degree to the north-east due to fluctuations in the MCMC fitting and are discounted. The remaining 30 candidates are listed in Table~\ref{tab:cands}. While the most likely candidate, \eso, is displaced to the $1\,\sigma$ contour, we note that the only other candidate host galaxy discussed by M18, WISEJ221621.59−191829.9 at z=0.024, is now excluded from the 3-sigma localisation region. As this method of localisation is not enough to uniquely identify a candidate galaxy, we proceed to consider other factors.

\begin{table*}
    \centering
    \begin{tabular}{c l c c c c c c c c l}
\hline
Rank & Galaxy & $p(s)$ & z & $p(z)$ & $n(z)$ & $p^\prime(z)$ & $m_r$ & $p(m_r)$ & $p(s) p\prime(z)$  & Other designations\\
(1) &(2) &(3) &(4) &(5) &(6) &(7) &(8) &(9) &(10) &(11) \\
\hline
1 & J221524.61-193504.8 & 0.046 & 0.00867 & 0.560 & 0.068 & 0.970 & 14.97 & 0.42 & 0.9845 & \eso \\
2 & J221427.49-200156.0 & 0.072 & 0.047 & 0.049 & 1.772 & 0.003 & 18.36 & 0.01 & 0.0052 & \\
3 & J221430.33-200013.6 & 0.088 & 0.049 & 0.037 & 2.343 & 0.002 & 17.55 & 0.02 & 0.0036 & \\
4 & J221338.59-193704.2 & 0.008 & 0.036 & 0.152 & 1.535 & 0.012 & 18.27 & 0.01 & 0.0022 & \\
5 & J221359.13-195106.0 & 0.125 & 0.055 & 0.011 & 2.741 & 0.000 & 17.66 & 0.01 & 0.0013 & \\
6 & J221445.43-194502.2 & 0.155 & 0.056 & 0.009 & 2.829 & 0.000 & 18.42 & 0.01 & 0.0013 & \\
7 & J221413.69-194032.1 & 0.106 & 0.056 & 0.009 & 2.838 & 0.000 & 18.34 & 0.01 & 0.0008 & \\
8 & J221548.31-192225.0 & 0.004 & 0.043 & 0.080 & 1.320 & 0.007 & 18.08 & 0.01 & 0.0007 & \\
9 & J221459.58-201030.4 & 0.010 & 0.056 & 0.010 & 2.756 & 0.000 & 18.65 & 0.00 & 0.0001 & \\
10 & J221601.96-193251.4 & 0.008 & 0.055 & 0.011 & 2.750 & 0.000 & 17.61 & 0.02 & 0.0001 & \\
11 & J221415.90-202202.4 & 0.001 & 0.046 & 0.056 & 2.005 & 0.003 & 18.43 & 0.01 & 0.0001 & \\
12 & J221501.93-200730.5 & 0.014 & 0.059 & 0.005 & 2.955 & 0.000 & 16.41 & 0.07 & 0.0001 & \\
13 & J221558.63-200327.8 & 0.004 & 0.057 & 0.008 & 2.852 & 0.000 & 18.19 & 0.01 & 0.0000 & \\
14 & J221404.53-195840.5 & 0.107 & 0.066 & 0.000 & 4.583 & 0.000 & 17.88 & 0.01 & 0.0000 & \\
15 & J221419.92-201529.1 & 0.007 & 0.063 & 0.001 & 2.527 & 0.000 & 16.42 & 0.07 & 0.0000 & \\
16 & J221358.74-193215.9 & 0.009 & 0.067 & 0.000 & 2.174 & 0.000 & 18.64 & 0.00 & 0.0000 & \\
17 & J221548.77-202123.1 & 0.001 & 0.063 & 0.001 & 4.186 & 0.000 & 17.22 & 0.02 & 0.0000 & \\
18 & J221437.97-192453.2 & 0.005 & 0.068 & 0.000 & 3.862 & 0.000 & 17.88 & 0.01 & 0.0000 & \\
19 & J221552.85-195808.4 & 0.009 & 0.070 & 0.000 & 3.447 & 0.000 & 18.03 & 0.01 & 0.0000 & \\
20 & J221559.40-192629.3 & 0.004 & 0.070 & 0.000 & 3.955 & 0.000 & 16.22 & 0.08 & 0.0000 & \\
21 & J221602.01-201021.0 & 0.001 & 0.066 & 0.000 & 4.588 & 0.000 & 18.07 & 0.01 & 0.0000 & \\
22 & J221449.49-192207.4 & 0.003 & 0.071 & 0.000 & 3.452 & 0.000 & 18.12 & 0.01 & 0.0000 & WISEA J221437.05-191904.9 \\
23 & J221357.33-195652.7 & 0.112 & 0.077 & 0.000 & 3.322 & 0.000 & 17.60 & 0.02 & 0.0000 & \\
24 & J221528.16-193851.9 & 0.055 & 0.076 & 0.000 & 3.306 & 0.000 & 18.45 & 0.01 & 0.0000 & \\
25 & J221458.56-185326.8 & 0.001 & 0.070 & 0.000 & 3.441 & 0.000 & 17.39 & 0.02 & 0.0000 & \\
26 & J221552.86-190559.2 & 0.001 & 0.072 & 0.000 & 3.721 & 0.000 & 18.52 & 0.01 & 0.0000 & \\
27 & J221645.10-195445.7 & 0.001 & 0.071 & 0.000 & 3.763 & 0.000 & 17.65 & 0.01 & 0.0000 & \\
28 & J221421.41-192756.0 & 0.008 & 0.076 & 0.000 & 3.311 & 0.000 & 17.71 & 0.01 & 0.0000 & \\
29 & J221416.08-193057.8 & 0.016 & 0.077 & 0.000 & 4.403 & 0.000 & 17.83 & 0.01 & 0.0000 & \\
30 & J221612.70-192222.1 & 0.001 & 0.076 & 0.000 & 3.309 & 0.000 & 17.47 & 0.02 & 0.0000 & \\
31 & J221503.26-192544.4 & 0.013 & 0.079 & 0.000 & 3.211 & 0.000 & 16.77 & 0.04 & 0.0000 & WISEA J221501.14-192536.9\\
32 & J221618.08-194206.2 & 0.006 & 0.079 & 0.000 & 3.207 & 0.000 & 16.93 & 0.03 & 0.0000 & \\
\hline
    \end{tabular}
    \caption{Candidate host galaxies for \frb. Column (1) indicates their rank, in order of most to least likely host (see labels on Figure~\ref{fig:lmap}); (2) \wise{} designation; (3) spatial localisation probability; (4) photometric redshift \citep[with the exception of the H\,{\sc i} redshift of galaxy 1, taken from][]{Meyer2004}, corrected for the North-South asymmetry in photometric $z$ derived from ANN$z$ \citep{WiseSuperCosmos}; (5) probability of that redshift according to the analysis of Section~\ref{sec:dm-z}; (6) number density of galaxies with that redshift per square degree per $0.001$ interval in $z$ in \wise{}; (7) redshift probability per galaxy, according to Equation~(\ref{eq:corrected_pz}); (8) $r$-band magnitude, corrected for North-South inconsistencies in $R$-band data due to the different passbands of the UKST in the South and POSS-II in the North \citep{WiseSuperCosmos}; (9) probability according to the PATH analysis of Section~\ref{sec:path}; (10) joint probability used to rank the candidates and normalised to unity over the candidates; and (11) apparent associations from the NASA/IPAC Extragalactic Database (NED).
    }
    \label{tab:cands}
\end{table*}

\subsection{Likely redshift of \frb}\label{sec:dm-z}
The DM of an FRB is a vital clue to its likely redshift due to the baryonic content of the Universe, as demonstrated by \citet{macquart_census_2020}. However, there is significant scatter about the ``Macquart relation'' between DM and $z$ --- due to different host galaxy contributions and inhomogeneities in the cosmic web --- leading to a broad distribution of DMs for a given redshift, $p({\rm DM}|z)$. However, $p({\rm DM}|z)$ has a robust lower bound due to the minimum density of voids and the (at least zero) contribution from host galaxies, leading to the reciprocal $p(z|{\rm DM})$ having a robust upper bound. Thus, the scatter in $p(z|{\rm DM})$ is smaller for low DMs. This probability is also a function of the properties of the detecting instrument, the intrinsic FRB population distribution and luminosity function \citep{James2022Letter}, and the signal-to-noise ratio (SNR) and time-width $w$ of the burst.

For \frb\,, we have accurate measurements of its DM (114.1 \ppcc) and SNR
(19.5) from \citet{shannon_dispersionbrightness_2018},
and a suitable limit on $w$ \citep[$<0.58$\,ms, effectively $w=0$; ][]{Qiu2020}, but no suitably reliable estimate of the detected beam position. These values do allow us to estimate $p(z|{\rm DM, SNR,} w)$. 
The measured SNR is important because bright FRBs are on-average closer than dimmer ones \citep{shannon_dispersionbrightness_2018}, while wider burst widths also require closer proximity to compensate for the increased noise over their duration.

We estimate $p(z|{\rm DM, SNR}, w)$ using the formulation from \citet{James2022Methods},
\begin{eqnarray}
p(z|{\rm DM, SNR}, w) & \propto &  \Phi(z) \frac{d V(z)}{d \Omega dz} p({\rm DM} |z)   \label{eq:pzgdmsw} \\
&&  \int dB \Omega(B) \frac{dp(s E_{\rm th}(B,w,{\rm DM},z))}{ds}, \nonumber
\end{eqnarray}
where $\Phi(z)$ is the FRB source evolution function; $V(z)$ is the comoving volume per solid angle per redshift interval per proper time; $p({\rm DM}|z)$ is the intrinsic probability distribution of dispersion measure as a function of redshift (including cosmological and host components); $\Omega(B)$ is the solid angle viewed at beam sensitivity $B$; and $p(s E_{\rm th}(B,w,{\rm DM},z))$ reflects the FRB luminosity function, i.e.\ it is the fraction of all FRB bursts from redshift $z$ emitted with energy above $s E_{\rm th}(B,w,{\rm DM},z)$, where $s \equiv {\rm SNR} / {\rm SNR_{\rm th}}$,
and $\rm SNR_{\rm th}$ is the FRB detection threshold. The energy threshold $E_{\rm th}$ is dependent upon $B$, $w$, DM, and $z$, where $w$ is the measured burst width. The constant of proportionality in Equation~(\ref{eq:pzgdmsw}) is the inverse of the expression integrated over redshift, which we leave out for brevity. We use best-fit parameter values from \citet{James2022Letter}, and refer readers to \citet{James2022Methods} for further details of the definitions of these functions.

\begin{figure}
     \centering
     \includegraphics[width=\columnwidth]{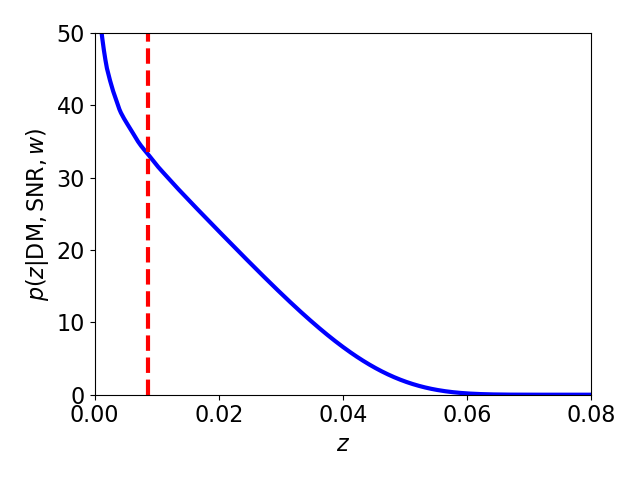}
     \caption{Likelihood of \frb\ coming from redshift $z$, given its observed DM, SNR, and width, for best-fit FRB population parameters from \citet{James2022Letter}. The red dashed line shows the redshift of \eso, $z$ = 0.00867.}
    \label{fig:pzgdmsw}
\end{figure}

Equation (\ref{eq:pzgdmsw}) is plotted in Figure~\ref{fig:pzgdmsw}, calculated for the best-fit FRB population parameters of \citet{James2022Letter}.
%, and a set of parameters lying within the 90\% confidence limits \citep{James2022Methods}. 
Since the best-fit slope of the luminosity function is relatively steep, a nearby host galaxy for \frb\ is strongly preferred.
%This is also the case for most models within the 90\% confidence limit (C.L.); only one, with the flattest possible intrinsic luminosity function, favours a host galaxy at higher redshift than \eso.% Both the best-fit model, and 8/12 within the 90\% C.L., show that $z=0.00867$ is at least twice as likely as $z=0.024$.

%\begin{figure}
%     \centering
%     \includegraphics[width=\columnwidth]{figures/114_p_z_g_dmsnr_ASKAP_FE.pdf}
%     \caption{Likelihood of FRB\,171020 coming from redshift $z$, given its observed DM, SNR, and width. Evaluations are for best-fit population parameters (blue solid), and parameters within the 90\% C.L.\ (grey dashed).}
%    \label{fig:pzgdmsw}
%\end{figure}

In order to use $p(z)$ as a Bayesian prior on the probability of a candidate galaxy being the true host, we take $p(z)$ as the probability of an FRB being detected from a given redshift range $z$ to $z +dz$. $p(z)$ is then the sum of the per galaxy prior probability, $p^\prime(z)$, over all $n(z)$ galaxies in that redshift range, i.e.\  $p(z) = \sum_i^{n(z)} p^\prime(z)$. Thus, any calculation of $p^\prime(z)$ should account for the increasing number of galaxies in the greater volume of the Universe probed in the interval $(z,z+dz)$ at larger redshifts.

To account for possible incompleteness in the \wise\ catalogue, we construct a histogram with linear bins of $dz$ and correct $p(z)$ by the number of galaxies in each bin $n(z)$, giving:
\begin{eqnarray}
p^\prime(z) & = & \frac{p(z)}{n(z)}. \label{eq:corrected_pz}
\end{eqnarray}
Table~\ref{tab:cands} lists these corrected probabilities $p^\prime(z)$, and the joint probability $p(s) p^\prime(z)$, which we use for ranking the candidates. We normalise all probabilities to sum to unity, which is equivalent to stating that the probability of the true host galaxy not being listed in the catalogues employed is zero. \eso\ is found to be the most likely host, with 98\% confidence. While other galaxies have spatial probabilities $p(s)$ up to four times higher than \eso, the probability of the host galaxy $p(z)$ lying at the redshift of \eso\ is at least four times higher than that from other galaxies. The greatest driver of the result however is the very low number density $n(z)$ of galaxies at \eso{}'s low redshift on the sky. At 0.68 per square degree, it would be an extraordinary coincidence for such a galaxy to be located in the localisation region purely by chance. We note however that this analysis does not consider the stellar shred ESO~601-G037 as a separate object, and cannot say anything about the relative likelihoods of it and \eso{} as host objects.

\subsection{Comparison with PATH methodology}\label{sec:path}
The Probabilistic Association of Transients to their Hosts \citep[`PATH';][]{path} methodology has recently been formulated and applied to the identification of the host galaxies of FRBs. It uses a Bayesian framework with the specific formulation incorporating priors on the $r$-band magnitude $m_r$ of the host galaxy, the offset distribution for an FRB from its putative host, and instrumental localisation uncertainties. 

Our analysis in Sections~\ref{sec:loc} \& \ref{sec:dm-z} is equivalent to a PATH analysis. Given that the instrumental localisation uncertainty is so large, the offset distribution of an FRB from its host can be ignored. The standard prior on the $r$-band magnitude distribution of FRB host galaxies used in PATH is a flat prior on an FRB originating in a galaxy with a given $m_r$, so that the prior for any individual galaxy is inversely proportional to the spatial number density of such galaxies on the sky. This method is the equivalent to our use of $p(z)$ normalised by the volumetric density of galaxies from the \wise{} catalogue.

It is interesting to compare the statistical power of our use of a prior on redshift $z$ compared to a prior on magnitude $m_r$ in this case. Taking the corrected r-band magnitudes from the \wise{} catalogue, we calculate the prior $p(m_r)$ according to the prescription of PATH, i.e.\ the priors are inversely proportional to the number densities from \citet{Diveretal2016_prior} and list the values in Table~\ref{tab:cands}. Using $p(m_r)$ alone clearly makes \eso\ the mostly likely host galaxy; however, the association only has a 42\% confidence, whereas using $p(z)$ alone yields a 97\% confidence in the association. We caution that hosts derived using $p(z)$ should not then be used as input for FRB population analyses since $p(z)$ is itself derived from such an analysis \citep{James2022Methods} and would result in a circular argument.

Ideally, $p(m_r)$ and $p(z)$ would be combined into a joint statistic since $p(z)$ alone ignores the fact that larger galaxies are more likely to be FRB hosts \citep{Bhandari2022}. As they are correlated however --- close galaxies are on-average brighter --- this would be a non-trivial exercise, which we leave to a future work.

\subsection{Limits on repetition} \label{sec:repetiton}
Our increased confidence in the association of \frb\ with \eso\ makes it the closest apparently non-repeating FRB. An extensive follow-up program with the {\em Murriyang} (Parkes) radio telescope and the Robert C.\ Byrd Green Bank Telescope (GBT) has failed to find any repetitions from the source, in addition to the $\sim$1150 hours that ASKAP has spent observing it, albeit mostly with a single antenna \citep{James2020repetitions}. Those authors
 use these non-detections to place a 90\% confidence level upper limit on the repetition rate of bursts above $10^{39}$\,erg in energy of 0.181--0.057 per day, assuming a differential power-law spectrum $dN(E)/dE \propto E^\gamma$, with $-0.7 \ge \gamma \ge -1.1$. However, this was a conservative upper limit calculated assuming a maximum redshift $z_{\rm max}$ of 0.0636.

The effective rate of visible bursts would be expected to scale (in a locally Euclidean Universe) as
\begin{eqnarray}
R & \propto & \left( \frac{z}{z_{\rm max}}\right)^{-2 \gamma}
\end{eqnarray}
where $z=0.008672$.
Thus our 90\% upper limits on the intrinsic rate of bursts will decrease by the same factor to 0.011--0.0007 bursts/day above $10^{39}$\,erg. By contrast, the rate of approximately one burst per hour observed from FRB~20121102A by the Five-hundred-meter Aperture Spherical radio Telescope (FAST; \citealt{FAST121102_bimodal}) corresponds to a rate of approximately six per day when accounting for the $\sim$25\% active duty cycle of that FRB \citep{121102_periodicity}. Accordingly, FRB~20121102A is at least 545 to 8450 times more active than \frb. Any model treating all FRBs as being intrinsically repeating sources must therefore account for at least 3--4 orders of magnitude difference in their rates.

\section{Multi-wavelength follow-up of \eso}\label{sec:follow-up}

\subsection{HI observations}
\label{sec:obs}

\begin{figure*}
	\includegraphics[width=0.95\textwidth]{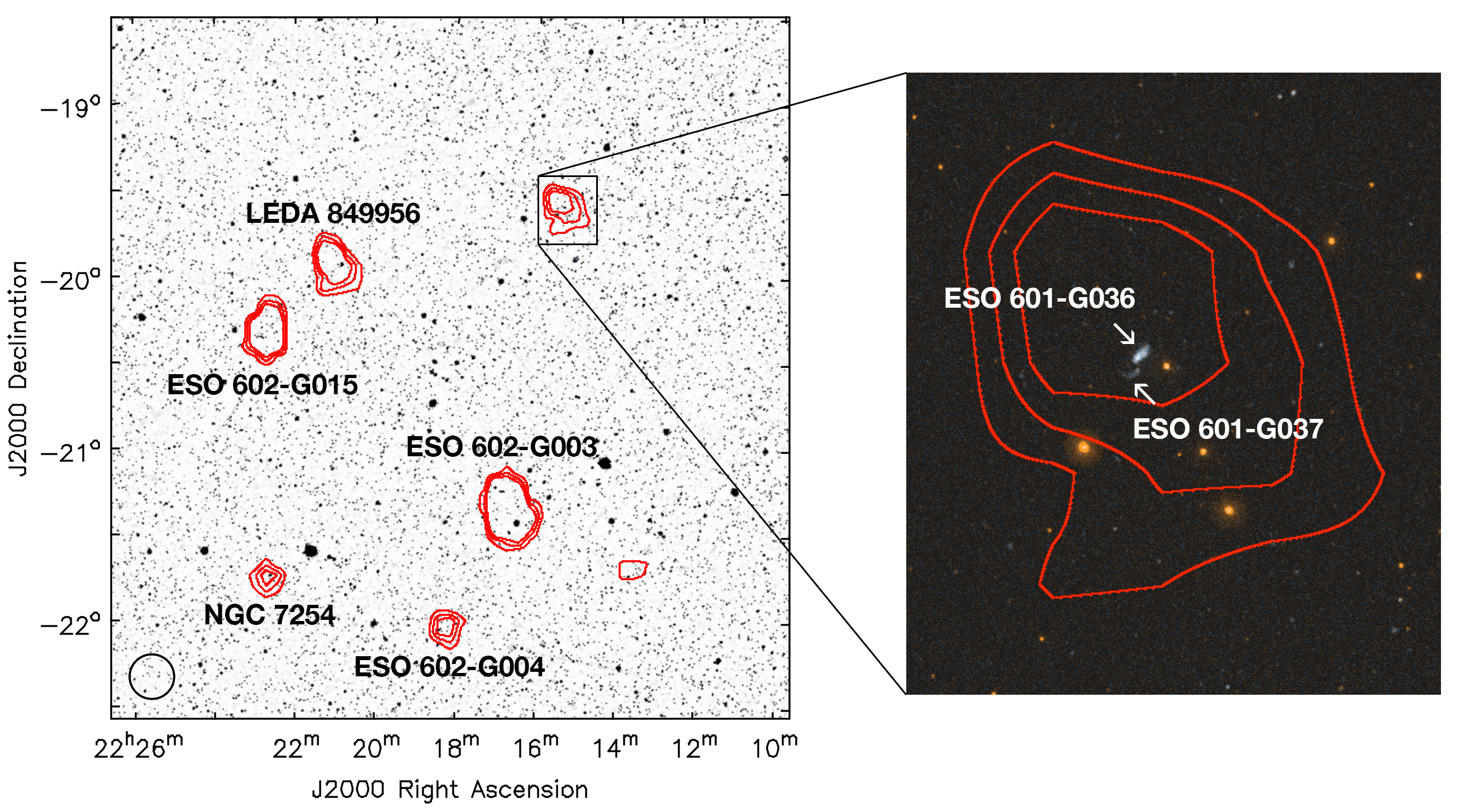}
    \caption{H{\sc{i}} total intensity contours from the HIPASS survey of \eso\ and neighbouring gas-rich galaxies superimposed on archival optical r-band Pan-STARRS1 \citep{chambers_pan-starrs} (left) and GALEX UV (inset at right) images. The HIPASS 15.5 arcmin beam is shown in the bottom left corner.}
    \label{fig:HIPASS}
\end{figure*}

Previous single-dish neutral hydrogen (\HI) observations from HIPASS (Figure \ref{fig:HIPASS}) show that the spatial extent of the gas in and around \eso\ and the nearby stellar shred ESO~601-G037 could easily fit within the Australia Telescope Compact Array (ATCA) field of view at 1.4~GHz with a single pointing.
In an attempt to resolve any \HI\ associated with ESO~601-G037 from that of \eso, we acquired 80~hours of observing time during semester 2019APR, under project code C3288. 
Preliminary test observations (project code CX417) indicated that the \HI\ component of the system was readily detectable by ATCA. 
Different array configurations were used to sample various spatial scales of the uv-plane, as summarised in Table \ref{tab:atcaobs}. 

\begin{table*}
	\centering
	\caption{ATCA observations of \eso.}
	\label{tab:atcaobs}
	\begin{tabular}{lccccc}
		\hline
Date & Project   & Array     & Integration &  \HI\ central frequency/ &Continuum central frequency/\\
        & code      & configuration   & time (h) &  bandwidth (MHz) &  bandwidth (MHz)\\
		\hline
13 Aug 2018 & C3211  & 1.5D      &  9.0  & --   & 5500/2048 and 9000/2048 \\
16 Aug 2018 & C3211  & H214      &  5.8  & --   & 16700/2048 and 21200/2048 \\
4 Nov 2018	& CX417	 & 750A	     &	2.4  & 1408/8.5   & 2100/2048 \\
2 Dec 2018	& CX417	 & H168      &	3.0  & 1408/8.5   & 2100/2048 \\
12 Dec 2018	& CX417	 & 1.5D      &	5.6  & 1408/8.5   & 2100/2048 \\
20 Apr 2019	& C3288	 & 750C      &	5.2  & 1408/8.5   & 2100/2048 \\
21 Apr 2019	& C3288  & 750C      &	6.8  & 1408/8.5   & 2100/2048 \\
 1 May 2019	& C3288	 & 1.5B      &	5.9  & 1408/8.5   & 2100/2048 \\
10 May 2019	& C3288	 & 1.5B      &	5.7  & 1408/8.5   & 2100/2048 \\
12 May 2019	& C3288  & 1.5B      &	5.6  & 1408/8.5   & 2100/2048 \\
15 May 2019	& C3288  & 1.5B      &	6.2  & 1408/8.5   & 2100/2048 \\
17 May 2019	& C3288  & 6A	     &	9.7  & 1408/8.5   & 2100/2048 \\
18 May 2019	& C3288  & 6A	     &	10.3 & 1408/8.5   & 2100/2048 \\
19 May 2019	& C3288  & 6A        &	7.1  & 1408/8.5   & 2100/2048 \\
24 May 2019	& C3288  & 6A	     &	9.1  & 1408/8.5   & 2100/2048 \\
\hline
\end{tabular}
\end{table*}

The observations used the ATCA Compact Array Broadband Backend \citep[CABB;][]{Wilson2011}, 1M-0.5k correlator mode, and two zoom bands. Each zoom band had a bandwidth of 8.5 MHz divided into 17409 channels, for a spectral resolution of 0.5 kHz. Both zoom bands were centred at 1408.25 MHz. This redundancy helped to mitigate the effects of possible hardware issues occurring in either band, such as periodic correlator block dropouts during observations that would stop data recording for a specific frequency range in the affected zoom band. 

Each observing run began with 10 minutes on the primary and bandpass calibrator PKS~1934-638, followed by a loop of alternating 3 minutes on the phase calibrator PKS~2135-209 and 40 minutes on the science target \eso. Every four loops, PKS~1934-638 was observed for another 5 minutes before continuing the loop. Any correlator block dropouts were noted and the alternate band was used for processing and imaging. 

The raw data were edited, calibrated, and imaged using \textsc{miriad} \citep{Sault1995} version 1.5 and its standard library of tasks. Each night of observations was processed individually, in a uniform manner, that incorporated visual inspection and manual outlier excision. During the imaging stage, the processed data were incorporated together to make each image cube. At this point, the 0.5~kHz channels were combined for a coarser resolution of 18.7~kHz (4~km~s$^{-1}$). Different weighting schemes and imaging parameters were tested to determine optimal settings for the intended science. 

For the most sensitive image cube we used natural weighting and did not include any data from ATCA antenna 6 (i.e.~using only baselines $<$3~km). The shortest baselines (with uv-range $<$ 0.25 k$\lambda$) had strong radio frequency interference (RFI) that caused a large scale ripple across the entire field of view and were also not used for the naturally-weighted image cube. For a higher spatial resolution cube, we included all the processed data and used a Briggs robustness of 0.5. The parameters for these two cubes are summarised in Table \ref{tab:process}. The root mean square (RMS) noise value for both cubes are comparable as the robust weighting incorporates additional baselines and is not as affected by the RFI in the shorter baselines. 

\begin{table}
	\centering
	 \begin{minipage}{\columnwidth}
	\caption{ATCA \HI\ processing and image cube details}
	\label{tab:process}
	\begin{tabular}{lcc}
		\hline
		                        & Natural    & Robust \\
		                        & weighting  & weighting \\
		\hline
		\textbf{Processing parameters}\\
		Antennas included       & 1--5            & 1--6 \\
		uv-range (k$\lambda$)   & $>0.25$         & 0.15--28.28 \\
        Image size (pixels)	    & 975 $\times$ 975 &2250 $\times$ 2250 \\
        Pixel size (arcsec)	    &4  	            &1.75\\
        Sidelobe suppression    &0                  &--\\
        Briggs robustness      &--                 &0.5\\
		\hline        
        \textbf{Image cube details}\\
        Beam size (arcsec)  &109 $\times$ 20    &44 $\times$ 9\\
        Spectral resolution (km s$^{-1}$) &4    &4\\
        Per channel RMS (mJy beam$^{-1}$) &0.8  &0.8\\
		\hline
	\end{tabular}
	\end{minipage}
\end{table}

\begin{figure*}
	\includegraphics[width=0.9\textwidth]{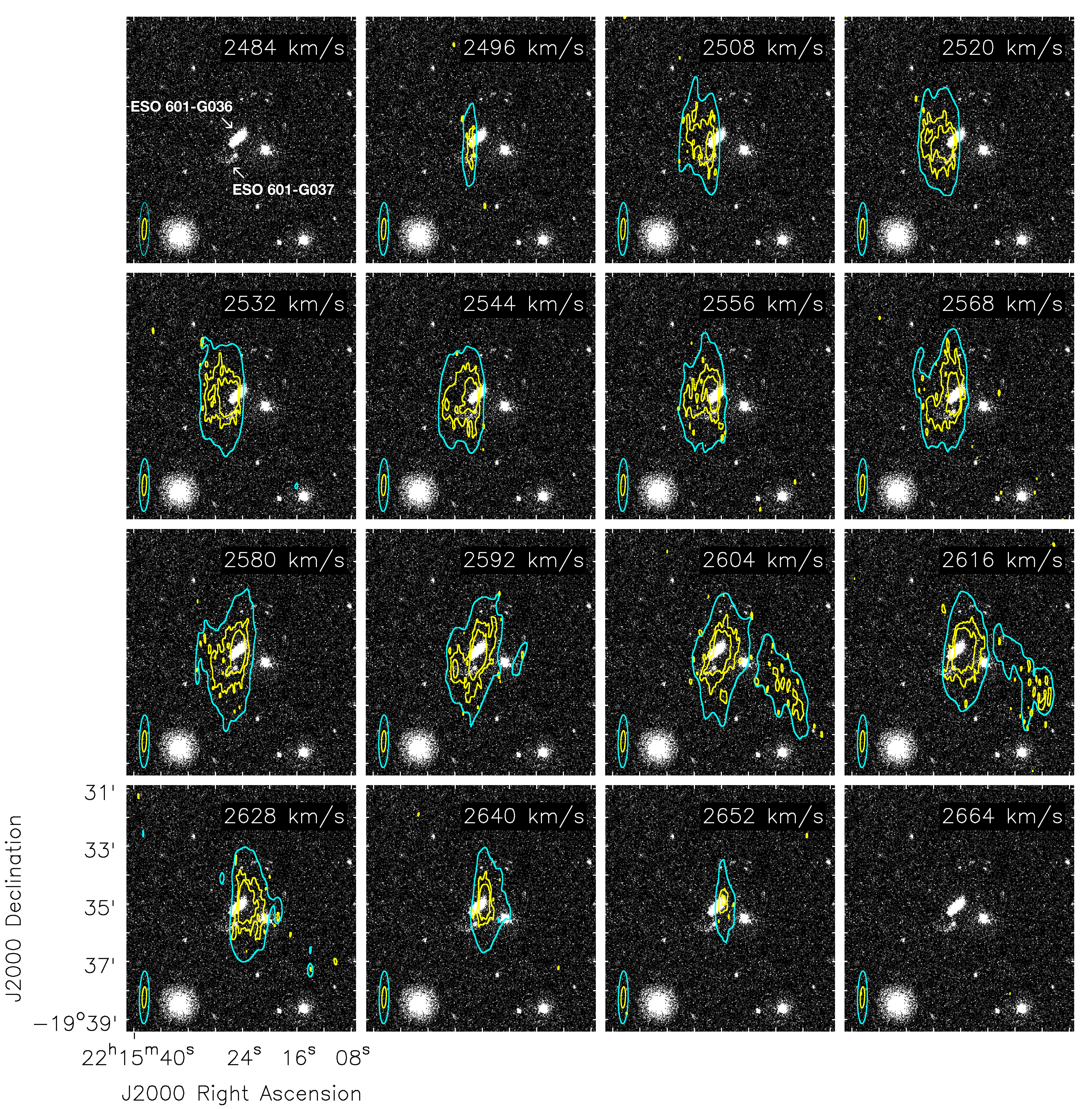}
    \caption{Channel maps of ATCA \HI\ cubes using a 3-channel average superimposed on a GALEX near-UV image. The cyan contour --- at 1.5 mJy beam$^{-1}$ --- is from the naturally-weighted cube with the gaseous tail clearly visible between 2604--2616 km s$^{-1}$. Yellow contours --- at 1.5, 3 mJy beam$^{-1}$ --- show the robust weighting. At least a portion of the \HI\ in the system, around 2568--2592 km s$^{-1}$, appears to follow the arc-like stellar structure of ESO 601-G037. The two different synthesized beams are shown in their respective colours in the bottom left corner of each panel.}
    \label{fig:chanmap}
\end{figure*}

\begin{figure*}
	\includegraphics[width=\textwidth]{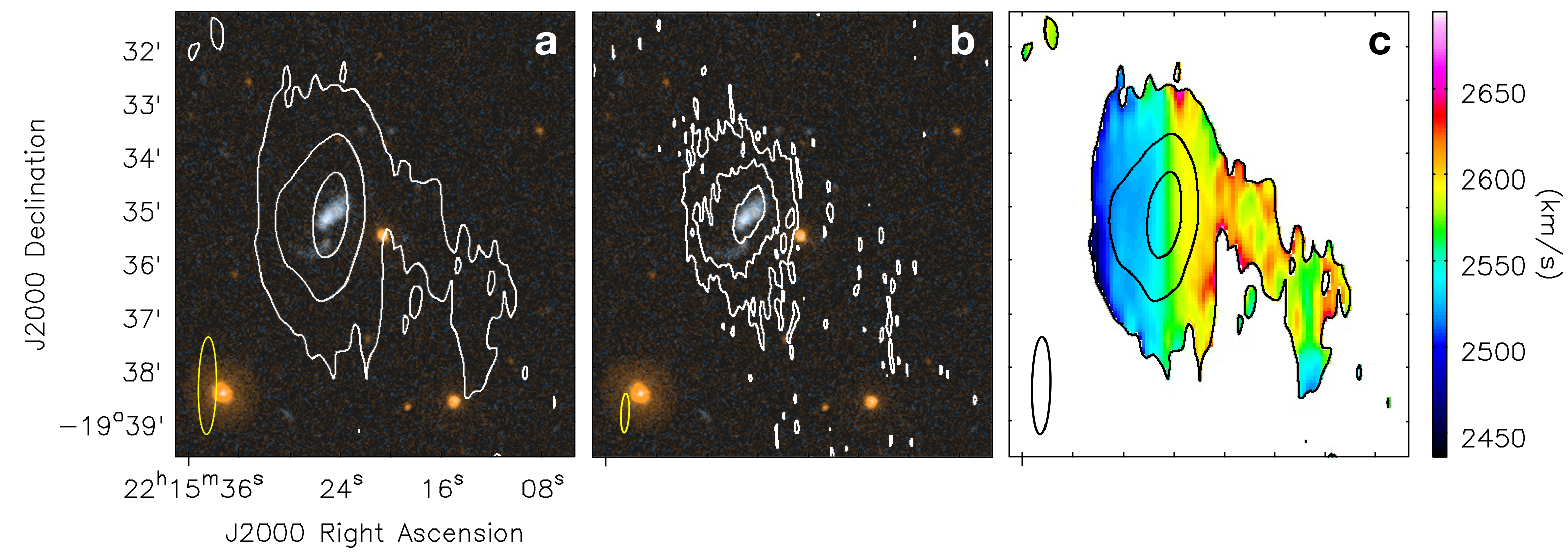}
    \caption{\HI\ moment maps of \eso\ showing a gaseous tail extending to the south-west of the stellar component. The synthesised beam is shown at lower left in each case. a) Total intensity \HI\ contours, with natural weighting, superimposed on a GALEX UV image. Contours are at (0.2, 2, 6) $\times$ $10^{20}$ atoms cm$^{-2}$. b) Total intensity \HI\ contours, with robust weighting, at (1.9, 5, 14) $\times$ $10^{20}$ atoms cm$^{-2}$. c) Velocity field map of the naturally-weighted cube.}
    \label{fig:moments}
\end{figure*}

 Figure \ref{fig:chanmap} shows channel maps of the system using a 3-channel average for conciseness. Both weighting schemes are shown as natural weighting captures the diffuse \HI\ and the robust weighting shows more detail of the higher density \HI\ peaks.
Moment maps showing the total \HI\ intensity of the system as well as the velocity field are presented in Figure~\ref{fig:moments}. The ATCA cubes have an elongated synthesised beam, due to the array's east-west configuration, which hindered clear separation of the \HI\ of \eso\ from ESO~601-G037. Nevertheless, a gaseous tail extending to the south-west of the stellar component of the system is clearly detected for the first time. The diffuse nature of the \HI\ tail is evident as it is barely detectable in the higher spatial resolution image. 

\begin{figure}
	\includegraphics[width=\columnwidth]{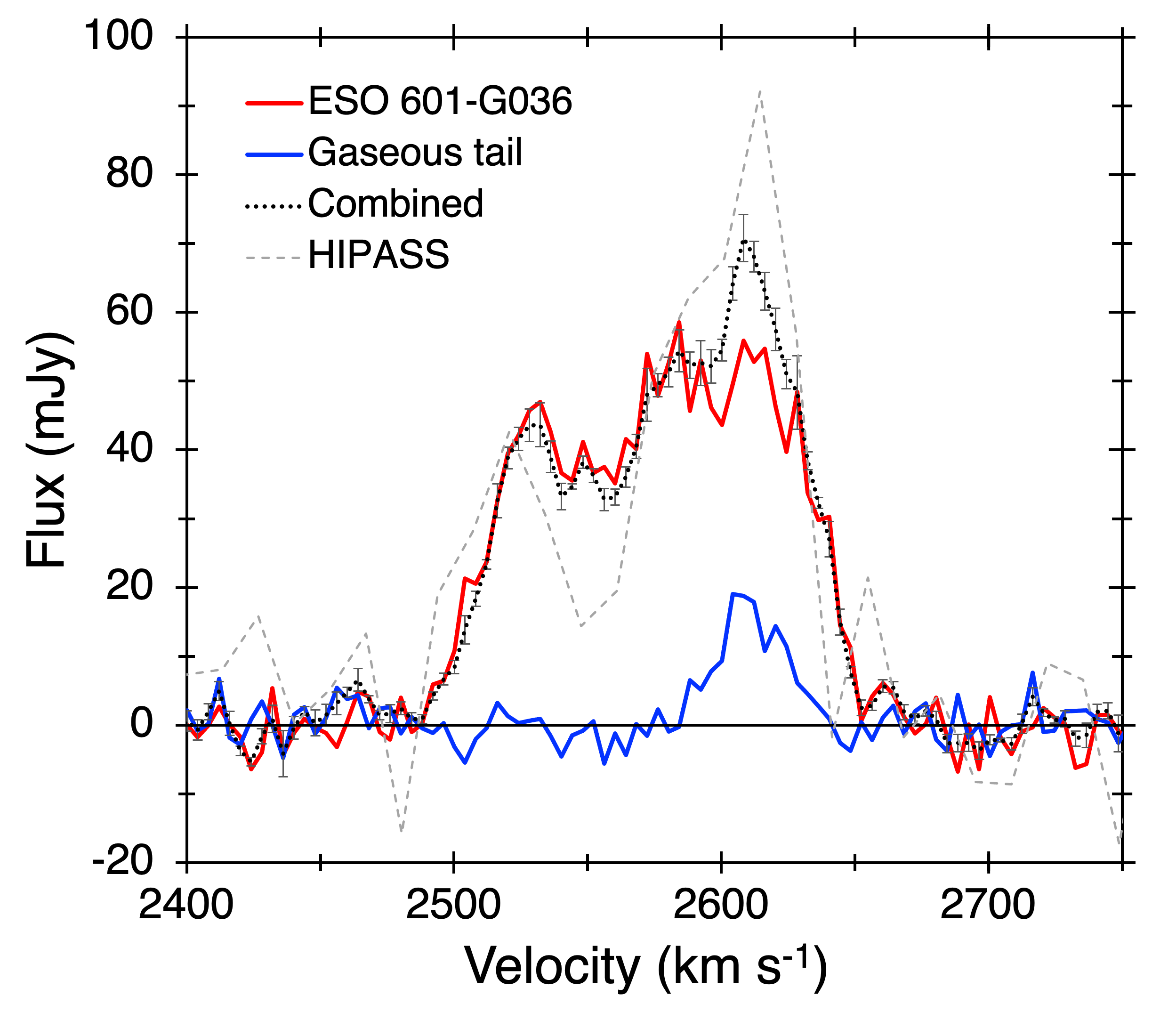}
    \caption{ATCA \HI\ spectra of \eso\ (and ESO~601-G037), the gaseous tail, and all sources combined. For comparison, the HIPASS spectrum --- which has higher sensitivity to diffuse emission --- has also been included.}
    \label{fig:spectra}
\end{figure}

A combined spectral profile incorporating all detected \HI\ within the immediate spatial and spectral vicinity of this system was measured both manually and using software tools such as \texttt{mbspect} in \textsc{miriad} and is presented in Figure~\ref{fig:spectra} along with profiles for \eso\ (combined with ESO~601-G037) and the gaseous tail. Overall, the combined ATCA spectrum is in good agreement with the slightly noisier, yet higher sensitivity HIPASS spectrum.

The spectra for the individual features were measured manually since there is no clear spatial or spectral separation between \eso\ and its \HI\ tail. As such, these plots represent an estimate of the flux contribution from each feature and no error bars have been included. Table \ref{tab:HI} summarises the \HI\ properties measured from the ATCA data. 

\begin{table*}
	\centering
	\caption{\HI\ properties measured from the ATCA observations. (1) Sources considered; (2) central \HI\ velocity; (3) \HI\ line width at 50\% of peak flux; (4) \HI\ line width at 20\% of peak flux; (5) peak \HI\ flux; (6) integrated \HI\ flux; and (7) \HI\ mass.
	}
	\label{tab:HI}
	\begin{tabular}{cccccccc}
	\hline
Source	&	v$_c$	&	$W_{50}$ 	&	$W_{20}$	&	Peak flux	&	$S_{\mbox{\HI}}$	&   $\log_{10}(M_{\mbox{\HI}} / M_\odot)$\\%M$_{HI}$    \\
	    &	(km s$^{-1}$)	&	(km s$^{-1}$)	&	(km s$^{-1}$)	&	(mJy)	&	(Jy km s$^{-1}$)	&    \\
(1)	    &(2)	&	(3)	&	(4)	&	(5)	&	(6)	&	(7) \\
\hline
\eso\	& 2578 $\pm$ 2		& 126 $\pm$ 2		& 148 $\pm$ 2		& 59 $\pm$ 4		& $\sim$ 5.2         &  $\sim$ 9.2\\%$\sim$ 1.7 \\
Gaseous tail    & 2613 $\pm$ 2		& 26 $\pm$ 2		& 48 $\pm$ 2		& 19 $\pm$ 1		& $\sim$ 0.5         &  $\sim$ 8.3\\%$\sim$ 0.2 \\
Combined   	    & 2576 $\pm$ 2		& 126 $\pm$ 2		& 148 $\pm$ 2		& 71 $\pm$ 3		& 6.4 $\pm$ 0.3   &  9.32 $\pm$ 0.02\\ %2.1 $\pm$ 0.1 \\
	\hline
	\end{tabular}
\end{table*}

\subsection{Radio continuum observations}
Broadband radio continuum observations of \eso\ spanning 2.1 to 21.2~GHz were made with ATCA as summarised in Table \ref{tab:atcaobs}. PKS~1934-638 was observed as the primary flux calibrator and the bandpass calibrator at frequencies below 10\,GHz. Observations at 16.7 and 21.2\,GHz central frequencies used PKS~1921-293 as the bandpass calibrator. PKS~2155-152 was used as the phase calibrator at all frequencies. 

The radio continuum data were also reduced following standard flagging, calibration, and imaging tasks in \textsc{miriad}. To retain sensitivity to the diffuse radio emission without heavily degrading spatial resolution, a Briggs robustness of 0.5 was used at 5.5 and 9.0\,GHz, and 0.0 at 2.1, 16.7 and 21.2\,GHz in the imaging step. 

\begin{figure}
    \centering
    \includegraphics[width=\columnwidth]{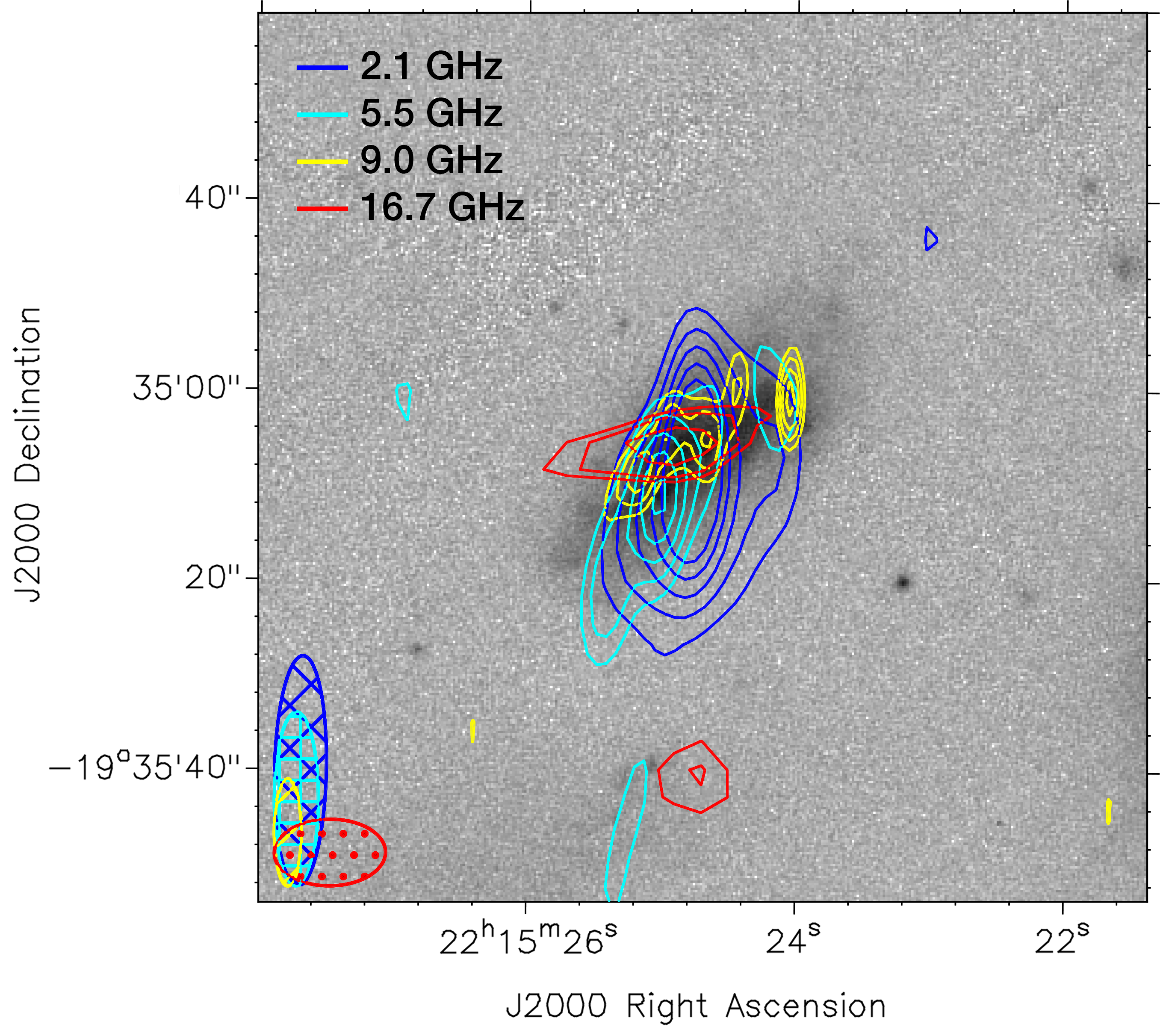}
    \caption{Optical r-band image of ESO 601-G036 from Pan-STARRS1 with 2.1\,GHz (blue; 70, 90, 110, 130, 150$\mu$Jy), 5.5\,GHz (cyan; 30, 40, 50, 60, 70$\mu$Jy), 9.0\,GHz (yellow; 20, 25, 30, 35$\mu$Jy) and 16.7\,GHz (red; 25, 30, 40$\mu$Jy) continuum contours overlaid. The synthesised beam from each frequency is shown in the bottom left corner. Most observations were carried out using a E-W array leading to an elongated beam in the N-S direction. The 16.7\,GHz data was observed with a hybrid array resulting in a different orientation of the beam.}
    \label{fig:cx_contours}
\end{figure}

An overlay of the radio emission detected at each frequency is shown in Figure \ref{fig:cx_contours}. Extended radio emission was detected for \eso\ at all frequencies up to 16.7\,GHz. No continuum emission was detected at 21.2\,GHz. There is a hint of emission within the spatial vicinity of ESO~601-G037; however, since it is brightest at 16.7 GHz and not detected at 9.0 GHz, the peaks are likely from noise in the continuum data.

To measure the flux density across the wide range of observed frequencies and synthesised beam sizes, images were first convolved to match the lowest resolution image at 2.1\,GHz. The flux density was then extracted at each frequency using the \textsc{miriad} task \texttt{imfit} over the same aperture defined by the source size at 2.1\,GHz (a Gaussian of 37.8 $\times$ 19.7~arcsec and position angle of $-18.3^{\circ}$). A summary of the measured flux densities along with the RMS and native resolution of each image are given in Table \ref{tab:radio}. A spectral index of $\alpha=-0.9$ is measured by fitting a simple power-law to the flux densities using the \texttt{scipy} package \citep{2020SciPy-NMeth}. This steep spectral index, combined with the clearly resolved nature of the continuum emission confirms that the radio emission is dominated by
synchrotron radiation from star formation processes, consistent with what was presented in M18.

\begin{table}
	\centering
	 \begin{minipage}{\columnwidth}
	\caption{ATCA radio continuum properties of ESO 601-G036. No continuum source is detected at 21.2\,GHz so a 3$\sigma$ upper limit is given.}
	\label{tab:radio}
    \centering
	\begin{tabular}{cccc}
		\hline
Frequency & resolution & RMS & Flux density \\
(GHz) & (arcsec) & ($\mu$Jy) & ($\mu$Jy) \\
\hline
2.1 & 24.2$\times$5.5 & 20.8 & 734.7 $\pm$ 5.4\\
5.5 & 18.3$\times$4.6 & 12.7 & 271.5 $\pm$ 2.7\\
9.0 & 11.1$\times$2.9 & 8.0 & 263.8 $\pm$ 1.6\\
16.7 & 11.7$\times$7.0 & 11.0 & 76.7 $\pm$ 2.8\\
21.2 & 9.4$\times$5.3 & 17.5 & $<$52.5\\
		\hline
	\end{tabular}
	\end{minipage}
\end{table}

\subsection{Optical imaging and photometry}

\begin{figure*}
	\includegraphics[width=0.9\textwidth]{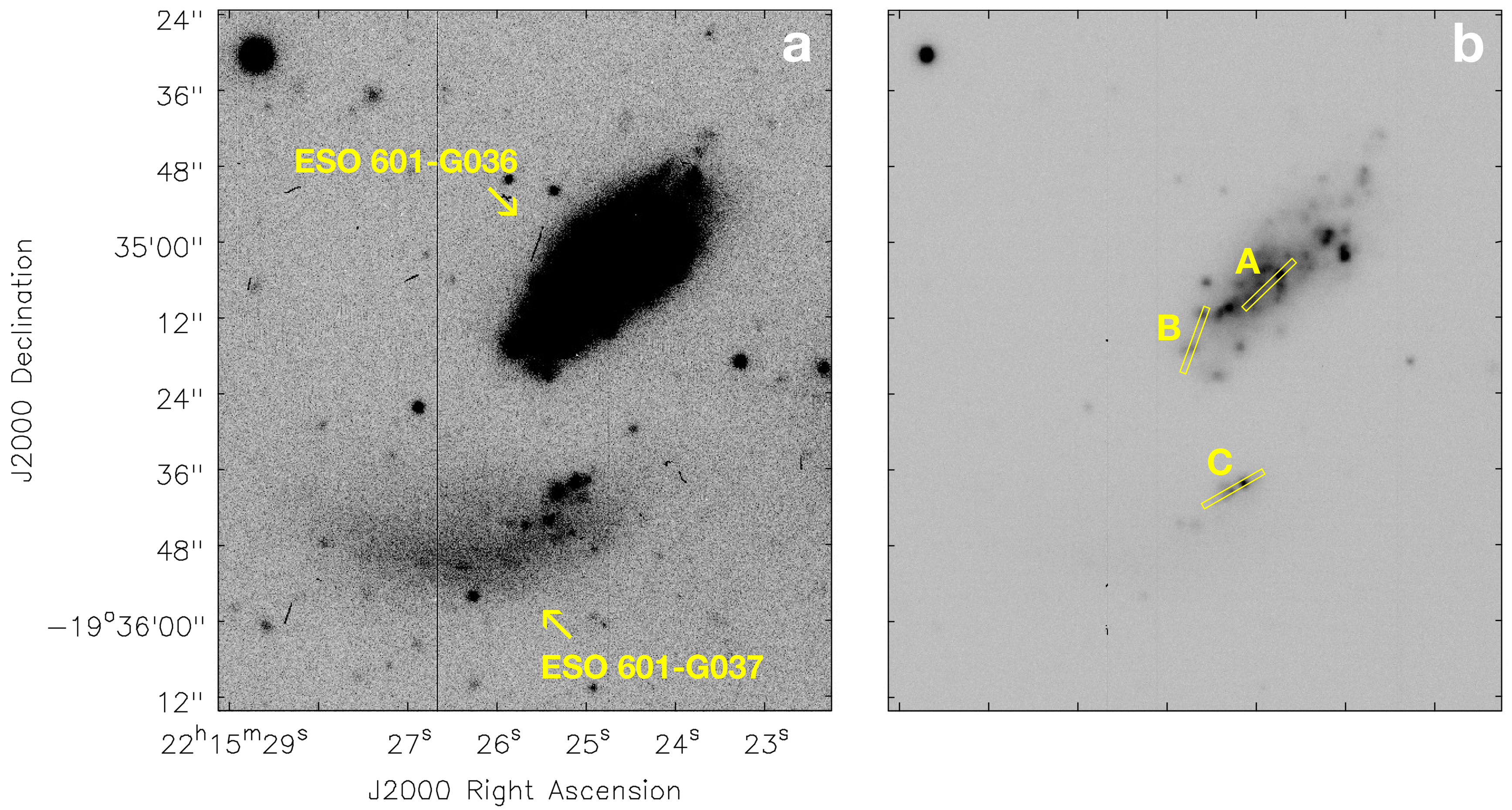}
    \caption{Gemini optical images of \eso\ and ESO~601-G037. a) r-band, scaled to bring out the arc-like morphology of ESO~601-G037. b) HaC, scaled to show the H\,{\sc ii}~region complexes. X-shooter slit positions for \eso\ (A and B) and ESO~601-G037 (C) are also shown. Each slit is 11\arcsec\ long by 1\arcsec\ wide.}
    \label{fig:gemini} 
\end{figure*}

After initial localisation of \frb, as part of a preliminary FRB follow-up strategy, optical imaging observations centred on \eso\ were taken on 15 July 2018 with Gemini-South (program GS-2018A-Q-205) using the GMOS HaC filter (6590--6650 \AA) and the broadband r-filter (5620--6980 \AA).  We obtained $3\times 180$\,s exposures with the HaC filter, and $1\times 180$\,s exposure with the r-filter. The data was processed using the Data Reduction for Astronomy from Gemini Observatory North and South (DRAGONS) package\footnote{\url{https://dragons.readthedocs.io}} using standard procedures.
The resulting images are shown in Figure \ref{fig:gemini} and show an arc-like structure of ESO~601-G037 that corresponds to the galaxy's shape revealed in the GALEX image (Figure \ref{fig:HIPASS}). The HaC imaging enable positioning of the X-shooter spectroscopy slits (see Section \ref{sec:spectroscopy}).  

The deeper Gemini r-band image was used as the reference image to determine flux-carrying pixels in the Pan-STARRS1 images \citep{chambers_pan-starrs}, taking advantage of data availability and self-consistent multi-band photometry provided by the latter dataset. After registering all images -- including the Gemini r-band image and the Pan-STARRS1 g, i, z, and y images -- to the Pan-STARRS1 r-band image, the detection mask and petrosain elliptical apertures were obtained from the Gemini image. Using the petrosain elliptical aperture, fluxes were derived for ESO 601-G036 from Pan-STARRS1 images and are presented in Table \ref{tab:photometry}. For ESO601-G037, the Gemini detection mask was used to derive fluxes. The different methods were used because \eso\ has a relatively regular galactic morphology, while ESO601-G037 is a distorted stellar shred. During this process, a standard photometric procedure following \citet{Wang2021} was utilised to conduct background subtraction, source finding, segmentation, and characterisation of photometry and morphological properties. The weight maps of Pan-STARRS1 were used to derive photometric uncertainties. Foreground extinction was corrected with the reddening map of \citet{Schlafly2011}. The (g-r) colour-based r-band stellar-mass-to-light ratio from \citet{Zibetti2009} was then used to estimate stellar masses, $M_\star$, which are also listed in Table \ref{tab:photometry}.

\begin{table}
	\centering
	 \begin{minipage}{\columnwidth}
	\caption{Stellar properties from Pan-STARRS1 photometry} 
	\label{tab:photometry}
    \centering
	\begin{tabular}{ccc}
		\hline
		                &\eso\    &ESO~601-G037\\
		\hline
        g-band mag	    &	15.382	$\pm$	0.003	&	17.66	$\pm$	0.01	\\
        r-band mag		&	15.141	$\pm$	0.002	&	17.49	$\pm$	0.01	\\
        i-band mag		&	15.102	$\pm$	0.002	&	17.31	$\pm$	0.01	\\
        z-band mag		&	15.06	$\pm$	0.01	&	17.33	$\pm$	0.02	\\
        y-band mag		&	14.88	$\pm$	0.01	&	17.23	$\pm$	0.06	\\
\hline									
$\log_{10}(M_\star / M_\odot)$	&	$\sim$8.5		&	$\sim$7.4	\\
		\hline
	\end{tabular}
	\end{minipage}
	\end{table}\label{sec:photo}

\subsection{Optical spectroscopy}

\begin{table}
	\centering
	\caption{{H\,{\sc ii}} region emission line properties measured from the X-shooter observations.}
	\label{tab:h2reg}
	\begin{tabular}{crrr}
	\hline
H\,{\sc ii} Complex	&	A	&	B 	&	C	\\
\hline
Aperture width ($^{\prime\prime}$) & 2.2 & 3.5 & 2.6 \\
\hline
~~ & \multicolumn{3}{c}{Dereddened line fluxes (H$\beta$=100)} \\
% ~~ & \multicolumn{3}{c}{(H$\beta$=100)} \\
\hline
[O\,{\sc ii}] $\lambda$3726    & 109.4 & 108.9  &  77.5 \\
{[O\,{\sc ii}]} $\lambda$3729  & 157.3 & 186.0  & 106.2 \\
H$\delta$ $\lambda$4101        &  27.7 &  27.7  &  27.2 \\
H$\gamma$ $\lambda$4340        &  50.9 &  50.0  &  48.6 \\
{[O\,{\sc iii}]} $\lambda$4363 &   5.2 & $<$2.7 &   9.8 \\
H$\beta$ $\lambda$4861         & 100.0 &  100.0 & 100.0 \\
{[O\,{\sc iii}]} $\lambda$4959 & 152.9 &   84.4 & 123.9 \\
{[O\,{\sc iii}]} $\lambda$5007 & 449.0 &  244.7 & 374.1 \\
{[N\,{\sc ii}]} $\lambda$6548  &   4.7 & $<$4.7 & $<$1.8 \\
H$\alpha$ $\lambda$6563        & 285.7 &  285.7 & 285.6 \\
{[N\,{\sc ii}]} $\lambda$6583  &  13.9 &   15.2 &   7.4 \\
{[S\,{\sc ii}]} $\lambda$6716  &  18.3 &   34.5 &  18.4 \\
{[S\,{\sc ii}]} $\lambda$6731  &  15.9 &   24.7 &   7.5 \\
\hline
$c$(H$\beta$)                  & 0.375 & 0.092  &  0.182 \\
$R_{23}$                       & 0.939 & 0.795   & 0.834 \\
N2O2                           & --1.282 & --1.287 & --1.397 \\
O32                            & 0.226  & --0.081  & 0.309 \\
$q$ ($10^{7}$ cm/s)            & 3--5   & 0.8--1.0 & 3.5--6 \\
12+log(O/H)                    & 8.45$\pm$0.05 & 8.43$\pm$0.02 & 8.23$\pm$0.05 \\
\hline
Mean redshift (km s$^{-1}$)           & 2558   &  2521 &  2587 \\
	\hline
	\end{tabular}
\end{table}

M18 used the X-shooter spectrograph \citep{X-shooter} mounted on UT2 (Kueyen) of the European Southern Observatory’s Very Large Telescope to effectively filter out other potential host galaxies, by virtue of the fact that only one was bright (and thus potentially close) enough to yield a redshift with just 5--6~minutes of exposure time. 
The same instrument was also used that night to obtain spectroscopy at the locations of three H\,{\sc ii}~region complexes in and around the \eso\ system, to probe the ionised gas phase kinematics and metallicity. Complex A lies close to the nucleus of \eso. Complex B is $\sim$10$\arcsec$ to the south-east along the major axis, near the edge of the optical disk. Complex C corresponds to the bright knots of star formation along the northern edge of ESO~601-G037, closest to \eso.
These locations are marked in Figure~\ref{fig:gemini}b. 

Slit widths of 1\farcs0 (UVB) or 0\farcs9 (VIS/NIR arms), and exposure times of 300~sec (VIS) or 360~sec (UVB/NIR arms) were used, but since diffuse gas emission tended to fill the 11$\arcsec$ slit length at these locations, matching observations of blank sky regions up to 35$\arcsec$ away were also obtained. Observations of the white dwarf EG~274 \citep{Hamuy1992} and the B9.5~V star HD~123247 were used for relative spectrophotometric calibration and telluric feature removal, respectively.

The echellograms for each X-shooter arm were processed using ESOReflex \citep{ESOReflex}, including debiasing, flatfielding, rectification, as well as wavelength and instrumental response calibration. 
Although automatic extraction of a point source from the centre of the 2D (slit position, wavelength) image is performed, this is not optimal for our purposes. 
Instead, the {\tt apall} task within V2.16.1 of {\sc iraf} was used to extract a spectrum of consistent spatial width in all arms about the H$\alpha$ peak, separately for each H\,{\sc ii}~Complex (Table~\ref{tab:h2reg}).
Gaussian fitting with the {\tt splot} task yielded the centroids and continuum-subtracted integrated fluxes of key emission lines. 
The fluxes have been corrected for interstellar reddening by assuming the intrinsic Balmer decrement for Case B recombination appropriate to a $T_{\rm e} = 10^4$~K H\,{\sc ii}~region, i.e. $I$(H$\alpha$)/$I$(H$\beta$) = 2.86, and the interstellar extinction curve of \citet{CCM1989}. The corrected fluxes are presented in Table~\ref{tab:h2reg}, along with the measured reddening parameter $c({\rm H}\beta)$ and mean velocity for each complex.

Traditionally the combined strength of the forbidden oxygen transitions can be used as an oxygen abundance diagnostic, via the $R_{23}$ parameter:
\vskip 3mm
\noindent
$R_{23}$ = ([O\,{\sc ii}] $\lambda\lambda3726,3729$ + [O\,{\sc iii}] $\lambda\lambda4959,5007$) / H$\beta$    
\vskip 3mm
\noindent
and its relationship to O/H.
However $R_{23}$ is known to be sensitive to both the ionisation parameter and to temperature, with the latter resulting in an ambiguity between an "upper" and a "lower" abundance branch \citep{KE2008}.
In order to break this ambiguity, the prescription laid out by \citet{SAMI2018} has been followed, which employs the N2O2 diagnostic:
\vskip 3mm
\noindent
N2O2 = log([N\,{\sc ii}]$\lambda6583$ / [O\,{\sc ii}]$\lambda\lambda3726,3729$)
\vskip 3mm
\noindent
and the criterion that for N2O2 $<-1.2$ (as is the case for all three complexes, Table~\ref{tab:h2reg}) the abundance falls on the "lower" $R_{23}$ branch.
Together with the value of the O32 diagnostic:
\vskip 3mm
\noindent
O32 = log([O\,{\sc iii}]$\lambda5007$ / [O\,{\sc ii}]$\lambda\lambda3726,3729$),
\vskip 3mm
\noindent
this initial metallicity estimate can place limits on the ionisation parameter $q$ using figure~5 of \citet{KK2004}. The oxygen abundance 12+log(O/H) is then given by the lower branch relation from eqn.~16 of \citet{KK2004}, with the uncertainty shown in Table~\ref{tab:h2reg} dominated by the range in $q$. These abundances are found to be consistent with the M18 estimations.\label{sec:spectroscopy}

\subsection{X-ray observations}

We observed \eso\ with the Neil Gehrels \textit{Swift} observatory \citep[\textit{Swift},][]{Gehrels2004} on 2021 May 19, for 1.7 ks (ObsID 00014317001). The observation was performed with the X-ray Telescope \citep[XRT,][]{Burrows05} in photon-counting mode and the Ultraviolet/Optical Telescope \citep[UVOT,][]{Roming05} with the UVW2 filter. The XRT data was reduced using the \texttt{xrtpipeline} v 0.13.6, \citep[\textsc{HEASoft} 6.29;][]{Blackburn1999, heasarc2014}. 

We do not detect any significant X-ray emission at the location of \eso. Using Poisson statistics for low-count experiments \citep[e.g.,][]{Gehrels1986, Kraft1991}, we calculate a 3$\sigma$ upper limit of $<$$6.8 \times 10^{-3}$ ct\,$\mathrm{s^{-1}}$ in the 0.3--10 keV band. Assuming a standard X-ray power law spectrum ($F\propto E^{-\alpha}$) with a photon index $\alpha = 2$ and the Galactic contribution to the hydrogen column density in the direction of the FRB, $N_{H,MW} = 1.9 \times 10^{20}~\mathrm{cm^{-2}}$ \citep{HI4PI2016}, we estimate an upper limit of $F_X \lesssim 2.5 \times 10^{-13}~\mathrm{erg~s^{-1}~cm^{−2}}$ on the unabsorbed X-ray flux in the 0.3--10.0 keV band. This corresponds to an upper limit of $L_X \lesssim 4.1 \times 10^{40}~\mathrm{erg~s^{-1}}$ on the X-ray luminosity. In comparison with the currently cataloged sample of ultra-luminous X-ray sources (ULXs), this limit is less than the observed luminosity of a small fraction of the brightest ULXs \citep{Kovlakas2020}, which have been proposed as FRB progenitors \citep{sridhar_periodic_2021,sridhar2022_ulx2}.

\subsection{Spectral energy distribution}

Using archival GALEX, Pan-STARRS1, VISTA, and WISE data, we fit the SED of \eso\ using the \textsc{ProSpect} SED fitting code \citep{RobothamProSpect2020}. 
We follow the implementation of \textsc{ProSpect} outlined in \cite{ThorneSEDFitting2021}. 
Briefly, we use the \cite{BruzualStellarpopulationsynthesis2003} stellar templates, assume a \cite{ChabrierGalacticStellarSubstellar2003} initial mass function, and model the dust attenuation and re-emission using the \cite{CharlotSimpleModelAbsorption2000} and \cite{DaleTwoParameterModelInfrared2014} models respectively while assuming energy balance. 
We adopt a skewed Normal star formation history parameterisation and an evolving metallicity history where the metallicity growth is mapped linearly to the stellar mass growth and the final metallicity is modelled as a free parameter.
We use the same priors as presented in table~2 of \cite{ThorneSEDFitting2021} and include a 10 per cent error floor across all bands to account for offsets between facilities and instruments.

\begin{figure}
    \centering
    \includegraphics[width = \columnwidth]{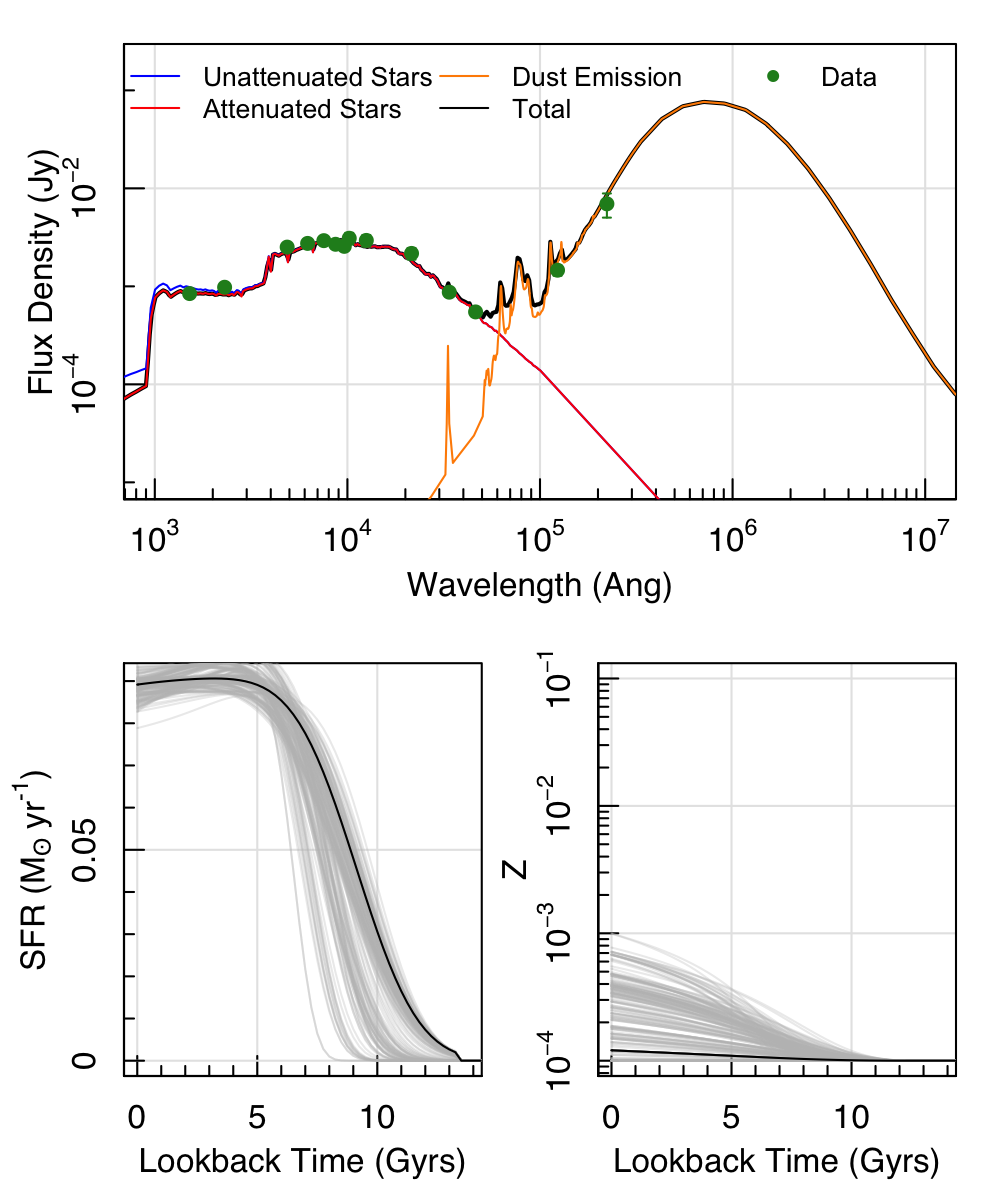}
    \caption{The upper panel shows the input photometry (green points) and resulting best-fitting SED (black), with the contribution from unattenuated stars (blue), attenuated stars (red), and dust emission (orange) shown. 
    The lower left panel shows the resulting best-fitting star formation history as the black line while the grey lines show the star formation histories of the rest of the posterior.
    The lower right panel shows the resulting metallicity history with lines as per the star formation history panel. 
    }
    \label{fig:OpticalSED}
\end{figure}

Figure~\ref{fig:OpticalSED} shows the input photometry, best-fitting SED, and resulting star formation and metallicity histories for \eso. 
From the \textsc{ProSpect} fit, we recover a stellar mass of $\log_{10}(M_\star / M_\odot) = 8.64^{+0.03}_{-0.15}$ (which is consistent to the mass calculated using the g-r colour method in Section \ref{sec:photo}) and a $\text{SFR} = 0.09 \pm 0.01\,M_\odot\,\text{yr}^{-1}$ (which is consistent to the previously reported value in M18). 
These values place \eso~on the star-forming main sequence at z$\sim$0, derived by \cite{ThorneSEDFitting2021}. 

In addition, we fit available GALEX and Pan-STARRS1 photometry for ESO 601-G037 using the same \textsc{ProSpect} implementation as described above.
ESO~601-G037 is undetected in the infrared imaging and so we limit our fitting to the available ultraviolet and optical data.  
We recover a stellar mass of $\log_{10}(M_\star / M_\odot) = 7.82^{+0.04}_{-0.8}$ (which is also comparable to the optical-only estimate) and a $\text{SFR} = 0.013 \pm 0.001\,M_\odot\,\text{yr}^{-1}$. The star formation history for ESO~601-G037 has a very similar shape to that of \eso\ (indicating constant star formation over the last $\sim$5 Gyr), with the former having a lower normalisation due to its lower mass.

\section{Discussion}\label{sec:discussion}
As described by M18, \frb\ has one of the smallest DMs ever detected and as such warranted further follow-up. Figure \ref{fig:lmap} shows the localisation area and after thorough analysis, we confirm M18's conclusion --- with 98\% confidence (see Table \ref{tab:cands}) --- that \eso\ was the host, which makes it the third closest FRB host galaxy. With an estimated upper limit on the repetition rate of $<0.011$\,bursts per day above $10^{39}$\,erg, \eso\ is also the closest host of an apparently non-repeating FRB. There is still a possibility that \frb\ originated from ESO~601-G037, but as we discuss in this section the two objects are connected and \eso\ is more dominant in the system.

\subsection{Signs of recent interaction}
The ATCA \HI\ observations have revealed a $\sim$3 arcmin long gaseous tail extending towards the south-west of \eso. This tail is fairly diffuse, contains $\sim$10\% of the {H\sc{i}} mass of the system, and spans a velocity width of $W_{20} = 48 \pm 2$ km s$^{-1}$ corresponding to the receding side of \eso\, (Figure \ref{fig:spectra}). 
As seen with other nearby FRB host galaxies \citep{Michalowski2021, Hsu2023}, the spectral profile of \eso\ is fairly asymmetric, indicating turbulent motion in and around the galaxy \citep{Glowacki2022}. It is likely that a disturbance has caused some gas from the eastern side of the galaxy to move towards the west, ultimately forming the \HI\ tail. Such disturbance could be an interloping galaxy --- with a central velocity closer to v$_c$ $\sim$ 2600 km s$^{-1}$ --- which may have also been stripped of at least some of its \HI\,, to contribute to the gaseous tail (see  \citealt{Bournaud2004}; \citealt{LeeWaddell2012}; \citealt{Lelli2015} for other examples). 
Considering other \HI\,-rich galaxies in the region are located over 1.4 deg ($\sim$900 kpc in projection) from \eso\ (see Figure \ref{fig:HIPASS}), the most obvious and possibly only interaction companion would be ESO~601-G037.

ESO~601-G037 has a pronounced arc-like stellar structure (see Figure \ref{fig:gemini}) that resembles a tidal tail (\citealt{Barnes1992}, \citealt{Paudel2018}). This stellar shred, located to the south of ESO 601-G036, is about an order of magnitude less massive but has a comparable specific SFR. The optical redshift obtained for ESO~601-G037 (Table \ref{tab:h2reg}, H\,{\sc ii} complex C) confirms it to be part of the \eso\ system, since it is within the velocity range observed for \eso\, and there does appear to be some \HI\ tracing the arc-like structure (see Figure~\ref{fig:chanmap}). ESO~601-G037's higher optical velocity (2587 km s$^{-1}$) is broadly consistent with the \HI\ tail ($v_c = 2613 \pm 2$ km s$^{-1}$ and $W_{20} = 48 \pm 2$ km s$^{-1}$). Furthermore, the oxygen line ratios of complex C are indicative of a less-enriched ISM, and hence a younger stellar population than complexes A and B in the main body of \eso, suggesting that star formation in ESO~601-G037 has been delayed relative to \eso. 

Overall, there are both kinematic and chemical arguments for ESO~601-G037 being a satellite companion that is undergoing a minor merger event with \eso, similar to the host system of FRB~20180916B \citep{Kaur2022}. The current spatial offset of ESO~601-G037 from the \HI\ tail (as much as $\sim$40~kpc) suggests that there has been at least one close passage affecting the \HI\ in the system and possibly another encounter transforming the stellar morphology of ESO 601-G037 \citep{Karera2022}. 

Figure \ref{fig:cx_contours} shows the radio continuum contours. Due to the elongation of the beam, the extension of the radio emission to the south of the galaxy may not be truly from the interaction between \eso\ and ESO 601-G037. Nevertheless, even accounting for the beam size, there does appear to be a general offset of the continuum emission from the optical disk of \eso, which is also consistent with the assumed interaction scenario.

\subsection{Other properties of \eso\,}

\eso\ is one of only a few FRB host galaxies that is detected in the radio continuum. The extended nature of the emission combined with the steep spectral index indicates that the emission is likely driven by star formation in the host galaxy, similar to what is seen is a small number of other FRB host galaxies (FRB~20191001A,  \citealt{bhandari_191001}; FRB~20190608B, \citealt{bhandari_host_2020}) and the host galaxy of the repeating FRB~20201124A \citep{fong_chronicling_2021, piro_20201124, ravi_2022}. We find no evidence for any compact, persistent radio source such as that detected in repeating FRBs FRB~20121102A \citep{marcote_repeating_2017} and FRB\,20190520B \citep{Niu2022_190520B}.

Using the measured spectral index, we calculate a 1.4~GHz radio luminosity for \eso\ of 1.89$\times$10$^{20}$\,W/Hz. Following the 1.4~GHz luminosity-to-SFR relation given in \citet{davies_2017}, we compute a SFR of 0.18$\substack{+2.54 \\ -0.17}$ $M_\odot\,\text{yr}^{-1}$. The relatively large uncertainities in the 1.4\,GHz-derived SFR are due to the scatter in the 1.4~GHz-SFR relation from \citet{davies_2017} and demonstrates that this method should only be used to provide an order-of-magnitude estimate of the SFR. Nonetheless, it is consistent with the SFR estimated from SED fitting, despite the methods probing different timescales; the 1.4\,GHz radio luminosity traces synchrotron emission from core-collapse SNe, whereas the SED fitting captures emission from hot young stars in the UV, right through to dust heating in the infrared by older, low-mass stars.

\subsection{Constraining an FRB progenitor model}

Minor mergers can cause increased star formation activity \citep{Hernquist1995,Lambas2012, Kaviraj2014}. \eso\ is clearly undergoing some form of interaction event, likely caused by the merging of ESO~601-G037, which aligns with the conclusions of other radio studies of FRB host galaxies that show evidence for recent interactions and/or significant spectral asymmetries \citep{Michalowski2021,Kaur2022,Hsu2023}. This finding is also consistent with FRB progenitor models predicting a tight temporal correlation with star formation; that is, emission from young magnetars formed from the collapse of massive stars. 

Our limits on repetition from \frb\ suggest that it was formed either from a cataclysmic merger, or is a significantly older object with correspondingly rarer and/or weaker emission. We have found that star-forming activity in \eso\ and ESO~601-G037 has been ongoing over the past $\sim$5 Gyr (lower left panel of Figure \ref{fig:OpticalSED}), which is consistent with both of these scenarios.

Nevertheless, recent \HI\ observations of the host galaxy of the non-repeating FRB~20211127I do not show evidence for any significant asymmetry \citep{Glowacki2023}. This particular host galaxy is similar in size and morphology to \eso, possibly suggesting that the two bursts could have a common origin mechanism despite their host galaxies having quite different interaction histories. A larger sample size of \HI\,-rich FRB host galaxies is required to draw any firm conclusions on the origin of FRBs in such systems.

\section{Conclusions}
We have used an updated localisation method and expectations for the DM--z distribution of FRBs, to search for the host galaxy of \frb\,.
We have confirmed, with 98\% confidence, that \eso\ is the host galaxy, as initially suggested by M18.
This result makes it, at $z=0.00867$ (37\,Mpc), the third closest confirmed FRB host galaxy and the closest host of an apparently non-repeating FRB. Our work strengthens previous limits on possible repetition of \frb\ to be less than 0.011 bursts/day above $10^{39}$\,erg.

The proximity of \eso\ and confirmation of it as an FRB host, motivated a series of follow-up observations.
We have carried out new \HI\ and radio continuum observations with ATCA, optical imaging with Gemini, optical spectroscopy with X-shooter on the VLT, and an X-ray search with {\em Swift}. We have also used archival data to measure stellar properties.

We find \eso\ to be a typical star-forming galaxy, with $\log_{10}(M_{\mbox{\HI}} / M_\odot) \sim 9.2$, $\log_{10}(M_\star / M_\odot) = 8.64^{+0.03}_{-0.15}$, and $\text{SFR} = 0.09 \pm 0.01\,M_\odot\,\text{yr}^{-1}$. Our \HI\ observations reveal the presence of a $\sim$3 arcmin long diffuse gaseous tail extending to the south-west of \eso\ that contains $\sim$10\% of the \HI\ mass of the system. Nearby stellar shred ESO~601-G037 is located to the south, has an arc-like morphology, SED-derived stellar mass of $\log_{10}(M_\star / M_\odot) = 7.82^{+0.04}_{-0.8}$, $\text{SFR} = 0.013 \pm 0.001\,M_\odot\,\text{yr}^{-1}$, and $\sim$0.2~dex lower gas metallicity than \eso.

We interpret the properties of this system as evidence of an ongoing interaction, likely the merging of ESO~601-G037 with the host galaxy, \eso. This finding is consistent with an FRB associated with recent star-forming activity, e.g.\ young magnetars. However, star formation has been ongoing for at least the last 5 Gyr and together with the strong limit on repetition, these properties are also consistent with the progenitor being a compact object merger. Our lack of an X-ray detection rules out the presence of the brightest ULXs, but is otherwise consistent with either FRB progenitor model.

Since the population of known FRB host galaxies is still small and the majority are much too distant for similar detailed follow-up observations that we have presented in this work, we encourage further analysis attempting to identify the host galaxies of nearby FRBs.

\section*{Acknowledgements}
%\small
We thank the reviewer for their comments and suggestions to improve the clarity of this paper. We also thank J. Xavier Prochaska and Marcin Glowacki for insightful discussions and suggestions that helped to improve this work.

%ATCA
The Australia Telescope Compact Array is part of the Australia Telescope National Facility (https://ror.org/05qajvd42) which is funded by the Australian Government for operation as a National Facility managed by CSIRO. We acknowledge the Gomeroi people as the traditional owners of the Observatory site.

%ASKAP
This scientific work uses data obtained from Inyarrimanha Ilgari Bundara / the CSIRO Murchison Radio-astronomy Observatory. We acknowledge the Wajarri Yamaji as the Traditional Owners and native title holders of the Observatory site. The Australian SKA Pathfinder is part of the Australia Telescope National Facility (https://ror.org/05qajvd42) which is managed by CSIRO. Operation of ASKAP is funded by the Australian Government with support from the National Collaborative Research Infrastructure Strategy. ASKAP uses the resources of the Pawsey Supercomputing Centre. Establishment of ASKAP, the Murchison Radio-astronomy Observatory and the Pawsey Supercomputing Centre are initiatives of the Australian Government, with support from the Government of Western Australia and the Science and Industry Endowment Fund.

% ESO & Gemini
This research is based on observations collected at the European Southern Observatory under ESO programme 0101.A-0455(B) (PI: J.-P. Macquart); as well as on observations obtained as part of program GS-2018A-Q-205 (PI: N. Tejos) at the international Gemini Observatory, a program of NSF’s NOIRLab, which is managed by the Association of Universities for Research in Astronomy (AURA) under a cooperative agreement with the National Science Foundation on behalf of the Gemini Observatory partnership: the National Science Foundation (United States), National Research Council (Canada), Agencia Nacional de Investigaci\'{o}n y Desarrollo (Chile), Ministerio de Ciencia, Tecnolog\'{i}a e Innovaci\'{o}n (Argentina), Minist\'{e}rio da Ci\^{e}ncia, Tecnologia, Inova\c{c}\~{o}es e Comunica\c{c}\~{o}es (Brazil), and Korea Astronomy and Space Science Institute (Republic of Korea). The Gemini data were processed using the DRAGONS (Data Reduction for Astronomy from Gemini Observatory North and South) package.

% NED
This research has made use of the NASA/IPAC Extragalactic Database (NED),
which is operated by the Jet Propulsion Laboratory, California Institute of Technology,
under contract with the National Aeronautics and Space Administration.

%CWJ
CWJ acknowledges support from the Australian Government through the Australian Research Council's Discovery Projects funding scheme (project DP210102103).

%Freya CSIRO scholarship
FONH was partially supported through a CSIRO postgraduate scholarship.

%Ryan
RMS acknowledges support through Australian Research Council Discovery (ARC) Project DP220102305.

\bibliography{ref}

\end{document}